\documentclass[twocolumn,eqsecnum,preprintnumbers,nofootinbib]{revtex4}
\usepackage{amsmath,amssymb,amsfonts,times,graphicx,color}

\newcommand{\beq}{\begin{equation}}
\newcommand{\eeq}{\end{equation}}
\def\bea{\begin{eqnarray}}
\def\eea{\end{eqnarray}}

\begin{document}

\title{Constructing holographic spacetimes using entanglement renormalization}
\author{Brian Swingle}
\email{brians@physics.harvard.edu}
\affiliation{Department of Physics, Harvard University, Cambridge MA 02138}

\date{\today}
\begin{abstract}
We elaborate on our earlier proposal connecting entanglement renormalization and holographic duality in which we argued that a tensor network can be reinterpreted as a kind of skeleton for an emergent holographic space.  Here we address the question of the large $N$ limit where on the holographic side the gravity theory becomes classical and a non-fluctuating smooth spacetime description emerges.  We show how a number of features of holographic duality in the large $N$ limit emerge naturally from entanglement renormalization, including a classical spacetime generated by entanglement, a sparse spectrum of operator dimensions, and phase transitions in mutual information.  We also address questions related to bulk locality below the AdS radius, holographic duals of weakly coupled large $N$ theories, Fermi surfaces in holography, and the holographic interpretation of branching MERA.  Some of our considerations are inspired by the idea of quantum expanders which are generalized quantum transformations that add a definite amount of entropy to most states.  Since we identify entanglement with geometry, we thus argue that classical spacetime may be built from quantum expanders (or something like them).
\end{abstract}

\maketitle

\section{Introduction}

A few years ago we proposed a connection between two new but superficially quite different approaches to quantum many-body physics, entanglement renormalization and holographic duality \cite{ent_ren_holo}.  Entanglement renormalization is motivated by considerations of entanglement in quantum matter while holographic duality emerges out of string theory, but both frameworks organize information in a quantum many-body system as a function of length scale according to the renormalization group.  After our initial proposal, a substantial body of work appeared supporting and extending the original idea \cite{tensor_geo_vidal,ent_ren_new1,ent_ren_new2,ent_ren_new3,cmera_holo_geo,ent_ren_cont,mera_branch,entren_follow1,entren_follow2,entren_follow3}.  Related work attempting to give formal holographic duals of generic field theories appeared in Refs. \cite{sslee_holo_qft1,sslee_holo_qft2,holo_qft_free}.  The connection between entanglement and geometry has also been studied in Ref. \cite{qg_ent_mark}.  In this paper we establish firmer contact between entanglement renormalization and holographic duality by considering entanglement renormalization in systems with many local degrees of freedom.   On the holographic side, many intriguing features emerge in the large $N$ limit including a classical geometry, entropy proportional to area, and a peculiar spectrum of operator scaling dimensions \cite{maldacena,witten,polyakov,holo_ee,holo_from_cft}.  It is our goal to show how all these features and more emerge naturally from entanglement renormalization.

Such a connection between entanglement renormalization and holographic duality is important.  Entanglement renormalization sheds light on the microstructure of holographic spacetime and gives us a guide to holographic duality away from large $N$ and strong coupling.  Thus we can learn about the physics of quantum gravity using tools of quantum information science and quantum many-body physics.  Holography, which appears to encode entanglement so naturally, can also teach us about quantum matter.  Indeed, two of the key problems of quantum matter are the understanding of many-body entanglement and the development of general entanglement based simulation tools.  We already know that entanglement renormalization can solve the one dimensional quantum Ising chain, so if it can also handle super-symmetric gauge theory in three dimensions then surely it can handle anything.  More practically, holography inspires new perspectives on entanglement renormalization including extensions to time dependent systems.  As these two subjects are brought closer together, the exchange should enrich both disciplines.

To begin we must understand the physical content of entanglement renormalization and holographic duality.  Entanglement renormalization \cite{vidal_er} is a real space renormalization group whereby quantum states, often ground states, are represented via a hierarchical tensor network structure.  This network geometry is motivated by the structure of entanglement in local quantum systems, with the physical picture being that we remove short range entanglement, coarse grain, and then repeat at a longer length scale.  We always only remove local entanglement at each step, so any long range entanglement is preserved deep into the infrared (IR) where it gives a clean characteristic of the quantum phase.  Entanglement renormalization can also be viewed as a variational state, known as the multi-scale entanglement renormalization ansatz (MERA) \cite{vidal_mera}, with a structure dictated by the removal of local entanglement followed by coarse-graining.  Such a formulation has a number of desirable features including efficient computation of correlators, efficient encoding of operator dimensions, and a simple representation of entanglement.

Holographic duality arose out of string theory \cite{maldacena,witten,polyakov} but is now understood as being considerably more general \cite{holo_ON,holo_from_cft}.  The duality states that an ordinary quantum field theory in $d+1$ dimensions is exactly equivalent to a theory of quantum gravity in a higher dimensional bulk geometry.  Every observable in the field theory has a corresponding representation in the gravity theory; for example, black holes in gravity are dual to finite temperature states in the field theory.  The best known example is the duality between $\mathcal{N}=4$ $SU(N)$ Yang-Mills theory in $3+1$ dimensions and type IIB string theory in an asymptotically $\text{AdS}_{5} \times S^5$ spacetime \cite{magoo}.  Ignoring the sphere (it is related to a large symmetry of the field theory), the extra holographic dimension in $\text{AdS}_{5}$ is related to renormalization group scale in the dual field theory.  Indeed, the ultraviolet (UV) of the field theory lives at the conformal boundary of AdS while the IR of the field theory lives deep inside the AdS space.

Our proposal is that these two frameworks, entanglement renormalization and holographic duality, are really two versions of the same basic structure.  In Ref. \cite{ent_ren_holo} we identified the emergent holographic direction with the number of coarse graining steps in entanglement renormalization.  We also showed how the resulting network describing entanglement renormalization could be interpreted as a discrete approximation to AdS where degree of quantum correlation, as measured by entanglement entropy $S = -\text{tr}(\rho \ln{(\rho)})$, was encoded geometrically in a metric $ds^2$.  In short, $ds^2=dS^2$ (literally true in $d=1$), so that the holographic geometry is built out of entanglement.  Our precise results included establishing a bound on the entropy in terms of discrete minimal curves, giving a geometric picture of correlation functions in terms of geodesics, and showing how black hole-like objects appeared at finite temperature.

In this paper we study entanglement renormalization in the context of large $N$ models, which are models with a large number of local degrees of freedom.  These include vector models like the $O(N)$ model \cite{ON_review} and matrix models based on a large $SU(N)$ gauge group \cite{magoo}.  We will primarily focus on the matrix model case, for reasons that will become clear as we go, but we will also discuss the situation with vector models.  Our primary goals include understanding the emergence of classical geometry and the peculiar spectrum of operator scaling dimensions in strongly coupled matrix theories \cite{holo_from_cft}.  However, we will also obtain a number of other salient features of the duality at large $N$ and strong coupling.

From the perspective of entanglement as geometry, roughly what we want is a quantum circuit or RG transformation that adds a definite amount of entanglement.  It turns out that there is already such a structure in quantum information science known as a quantum expander \cite{qexpander1,qexpander2}.  Hence we want to argue that classical space is made of quantum expanders (or something like them).  A quantum expander may be understood as a generalized transformation of a quantum state (density matrix) that adds a definite amount of entropy for almost all inputs.  Although we will not make precise use of the quantum information notion, it is an important motivation which we review below.

Now for a preview of our main results.  We argue that, at large $N$ and strong coupling, the tensors of the tensor network will be strongly mixing which implies both the addition of a definite entropy and the largeness of most operator dimensions.  In this context, we point out the potential relevance of quantum expanders and related notions to spacetime geometry.  We also derive the phase transition structure of holographic mutual information from entanglement renormalization, and furthermore, we show how to compute sub-leading corrections.  We show how to obtain various $n$-point functions from the geometric structure of entanglement renormalization and point out a similarity to Witten diagrams.  We also give a holographic interpretation of the branching MERA and a new perspective on Fermi surfaces that should have holographic ramifications.  Finally, we recast our considerations in the context of continuous entanglement renormalization, discuss the idea of RG causality, elaborate on our black hole-like objects, and obtain recent holographic metrics with logarithmic entanglement entropy from a field theory construction.

This is a long paper, so we indicate where the main results are located.  Sec. II is the toolbox where we review certain key ideas including conformal symmetry, large $N$ field theories, entanglement and entanglement renormalization, and the idea of quantum expanders.  Only the entanglement renormalization section is really crucial for what comes after.  Sec. III contains the first of our main arguments in the context of large $N$ entanglement renormalization on a lattice. There we discuss operator dimensions, the emergence of area, the structure of mutual information and correlations, the idea of RG causality, and some ideas on Fermi surfaces.  Sec. IV is a shorter section in which we compare the results of Sec. III to holography proper.  In Sec. V we reformulate our results in the context of continuous MERA.  There we discuss entanglement and correlations, the idea of RG causality, black hole-like objects, the idea of entanglement per scale, and a field theory construction of holographic metrics supporting area law violations.  Finally, we conclude the paper with a discussion of numerous open issues and future directions.

\section{Toolbox}
Here we briefly review some background information necessary for the arguments in the paper.  We begin with scale invariance in quantum field theory with a special emphasis on conformal symmetry.  This symmetry is important because a critical part of our discussion concerns primary (scaling) operators and their scaling dimensions.  Then we discuss the large $N$ limit in field theory and describe some of the systems of relevance to us.  Next we review entanglement renormalization and the proposal in Ref. \cite{ent_ren_holo}.  Finally, we review the notion of quantum expanders and random unitaries and briefly discuss their role in inspiring our ideas.  Readers familiar with some of this material are encouraged to skip forward to the new arguments in Sec. III.

\subsection{Conformal symmetry}
We will be primarily interested in scale invariant theories.  In the relativistic case these often turn out to also be conformally invariant (see Ref. \cite{condmat_holo_review} for a friendly introduction).  We take the metric of the spacetime in which the field theory lives to have $g_{tt} < 0$.  For example, in two dimensional flat space we would have $ds^2 = g_{tt} dt^2 + g_{xx} dx^2$ with $g_{tt} = -1$ and $g_{xx}=1$.  Conformal transformations as those that preserve the form of the metric up to rescaling and are a generalization of rotations, translations, and scaling transformations or dilatations. $d$ always refers to the spatial dimension of the field theory.

The conformal group is generated by the momentum (translations),
\beq
P_\mu = - i \partial_\mu,
\eeq
the angular momentum (rotations/boosts),
\beq
M_{\mu \nu} = i(x_\mu \partial_\nu - x_\nu \partial_\mu),
\eeq
the dilatation generator (dilatations),
\beq
D = - i x^\mu \partial_\mu,
\eeq
and the special conformal generators (special conformal transformations),
\beq
K_\mu = - i (x\cdot x \partial_\mu - x_\mu x \cdot \partial).
\eeq
We can easily check, for example, that the translation operator $\mathcal{T}(a) = e^{i a P}$ changes $x$ as
\beq
\mathcal{T}(a) x^\mu \mathcal{T}^{-1}(a) = e^{a^\nu \partial_\nu} x^\mu e^{ - a^\nu \partial_\nu} = x^\mu + a^\mu
\eeq
while the dilatation generator gives
\beq
e^{i b D} x^\mu e^{-i b D} = b x^\mu.
\eeq

More generally, in a field theory with conformal symmetry (CFT) we can relate all these generators to the stress tensor $T_{\mu \nu}$ as follows:
\bea
&& P^\mu = \int d^d x \,T^{\mu t}, \cr \nonumber \\
&& M^{\mu \nu} = - \int d^d x \left( x^\mu T^{\nu t} - x^{\nu} T^{\mu t} \right), \cr \nonumber \\
&& D = \int d^d x \left( x_\mu T^{\mu t}\right), \cr \nonumber \\
&& K^\mu = \int d^d x \left(x \cdot x T^{\mu t} - x^\mu x_\nu T^{\nu t}\right).
\eea
The expression for $P^\mu$ gives insight into the conventional physics of the stress tensor.  $T^{tt}$ or $T^{00}$ is the energy density while $T^{i0}$ is the momentum density in the $x^i$-direction.  Related to each generator is a conserved current.  For example, the dilatation current is $J_D^\mu = x_\nu T^{\nu \mu}$ so that
\beq
D = \int d^d x J_D^t
\eeq
and
\beq
\partial_\mu J^\mu_D = g_{\mu \nu} T^{\mu \nu} + x_\nu \partial_\mu T^{\nu \mu} = 0
\eeq
where the last equality follows because the stress tensor is conserved and traceless (in a CFT).

Besides the usual Lorentz algebra, the important new commutation relations involve the dilatation and special conformal generators.  These relations include
\bea \label{cftalg}
&& [D,P_\mu] = i P_\mu, \cr \nonumber \\
&& [D,K_\mu] = - i K_\mu, \cr \nonumber \\
&& [K_\mu, P_\nu ] = 2i ( g_{\mu \nu} D - M_{\mu \nu}),
\eea
in addition to the statement that $D$ is a scalar and $K_\mu$ is a vector under rotations.  Since the momentum has scaling (mass) dimension one, we expect that an operator $O$ with definite scaling dimension $\Delta$ will satisfy $[D,O] = i \Delta O$.  The $[D,K_\mu]$ commutator then implies that $K_\mu$ lowers the dimension of any operator on which is acts, that is
\bea
&& [D,[K_\mu, O]] = - [K_\mu,[O,D]] - [O,[D,K_\mu]] \cr \nonumber \\
&& = - [K_\mu , -i \Delta O] - [O,-i K_\mu] \cr \nonumber \\
&& = i (\Delta - 1) [K_\mu,O].
\eea
Since the scaling dimension of sensible operators must be positive we see that the special conformal transformations must annihilate some fields, called primary fields.

As an aside, special conformal transformations may be unfamiliar to some readers.  They can be understood as the following three step process.  First, we invert all coordinates
\beq
x^\mu \rightarrow \frac{x^\mu}{x^2},
\eeq
then we translate the inverted coordinates
\beq
\frac{x^\mu}{x^2} \rightarrow \frac{x^\mu}{x^2} + b^\mu,
\eeq
and finally we invert the coordinates again
\beq
\frac{x^\mu}{x^2} + b^\mu \rightarrow \frac{\frac{x^\mu}{x^2} + b^\mu}{x^{-2} + 2 b\cdot x/x^2 + b^2} = \frac{x^\mu + x^2 b^\mu}{1 + 2 b\cdot x + b^2 x^2}.
\eeq
Although we tend to focus on simple scaling transformations, these more complex special conformal transformations also provide important constraints on correlation functions.  Additionally, they play an important role in the entanglement entropy of spheres in CFTs \cite{holo_ee_sphere}.

Finally, let us mention the concept of an operator product expansion.  This permits us to make sense of products of scaling operators in the continuum theory.  When two scaling operators are brought close together, a singularity will develop which is cutoff by the lattice spacing.  In the continuum we may write
\beq \label{ope}
O_i(x) O_j(y) \sim \sum_k c_{ij}^k(x-y) O_k(y)
\eeq
where the operator product expansion (OPE) coefficients $c_{ij}^k$ are singular as $x$ approaches $y$.  If we suppress spin labels then the OPE coefficients have the form
\beq
c_{ij}^k(x-y) \equiv c_{ij}^k \frac{1}{|x-y|^{\Delta_i + \Delta_j - \Delta_k}}
\eeq
which is indeed singular for small enough $\Delta_k$.  The meaning of this expansion is that it becomes an equality when inserted into correlation functions with no other nearby operators.  The fields $O_k$ appearing on the right hand side are called fusion products of the fusion of $O_i$ and $O_j$.  The OPE coefficient for the identity $O_k = 1$ is simply the two-point of the two operators.  More generally, the full set of OPE coefficients and operator dimensions are a crucial part of the data defining a conformal field theory (see for example Ref. \cite{holo_from_cft}).

\subsection{Large $N$ limit}

The simplest example of a large $N$ theory is provided by the vector model with $O(N_f)$ symmetry.  Here $N=N_f$ and the basic fields are $N_f$ component vectors $\phi_a$.  The simplest Lagrangian incorporating this field is provided by the free theory
\beq
\mathcal{L} = \frac{1}{2}\left((\partial \vec{\phi})^2 - m^2 (\vec{\phi})^2\right).
\eeq
Correlators obey $\langle \phi_a \phi_b \rangle = \delta_{ab}$, so the theory is equivalent to $N$ copies of a free scalar field.  We can add interactions to $\mathcal{L}$ via a quartic term as follows
\beq
\mathcal{L}  = \frac{1}{2}\left((\partial \vec{\phi})^2 - m^2 (\vec{\phi})^2\right) - g_4 ((\vec{\phi})^2)^2.
\eeq
Note however that the last term contains roughly $N^2$ terms while the first terms both contain $N$ terms, so we see that to keep the action of order $N$ as $N\rightarrow \infty$ $g_4$ must be chosen so that $\tilde{g}_4 = N g_4$ has a finite limit.  In this limit, the whole action becomes large and the theory is solvable in a saddle point approximation.  It will be important later that the number of low dimension primary fields diverge as $N\rightarrow \infty$ i.e. $\phi_a$ is primary for all $a=1,...,N$.

Many interesting phenomena are visible in the limit $N\rightarrow \infty$, including symmetry breaking and a non-mean field transition into the symmetry broken phase \cite{ON_review}.  The quantum phase transition to the symmetry broken state occurs at a particular value of $\tilde{g}_4$ and is described by an interacting conformal field theory.  Nevertheless, the spectrum of primary operators in the large $N$ limit looks much like the spectrum of the free theory in the massless limit $m^2 =0$.  Finally, at the level of a microscopic lattice model, there are numerous simple rotor and spin models that have the above theory as a low energy limit.

A more interesting kind of large $N$ limit is provided by gauged matrix models \cite{magoo}.  In these models we consider a gauge group like $SU(N_c)$ and consider theories of gauge fields $A$ coupled to matter in various representations of the gauge group.  The gauge field itself is in the adjoint representation of the gauge group, which means that it may be understood as an $N_c \times N_c$ matrix.  Similarly, the matter fields $\Phi$ may also be in the adjoint representation in which case they are also $N_c \times N_c$ matrices.  To construct gauge invariant operators, we need only multiply such matrix operators and take a trace since adjoint fields transform like $\Phi \rightarrow U_c \Phi U_c^\dagger$ with $U_c \in SU(N_c)$. These gauge models, which are referred to as matrix models, have $N=N_c^2$ and are the most directly relevant to holographic duality.  Typical operators in these models are traces (over the matrix degrees of freedom) of products of adjoint fields e.g. $\text{tr}(\Phi^2)$ or $\text{tr}(F^2)$ where $F \sim dA + [A,A]$ is the field strength of $A$.

A typical Lagrangian for a matrix model takes the form
\beq
\mathcal{L} = \frac{1}{g^2} \text{tr}\left((D\Phi)^2 + m^2 \Phi^2 +  \Phi^3 + \Phi^4 + ...\right)
\eeq
in terms of a matrix field $\Phi$ and its covariant derivative $D\Phi = \partial \Phi + [A,\Phi]$ \cite{condmat_holo_review}.  If we rescale $\Phi/g \rightarrow \Phi$ it becomes clear that $g$ is a coupling constant.  In this case we have $N\sim N^2_c$ and thus we must again tune $g$ to zero as as $N$ gets large to have a sensible theory.  The correct large $N$ coupling turns out to be $\lambda = g^2 N_c = g^2 N^{1/2}$ which is called the 't Hooft coupling.  It would take us too far afield to properly derive this result, so we restrict ourselves to a few comments.

First, in the large $N$ limit perturbation theory simplifies so that only so-called planar diagrams survive.  The planar diagrams are then organized into a perturbation series in powers of $\lambda$ with all other non-planar diagrams suppressed by powers of $1/N$.  Second, one simple way to understand the scaling of $\lambda$ is to look at the RG $\beta$ function of the coupling.  Typically we have
\beq
\partial_u g = b N_c g^3 + ...
\eeq
with $b \sim 1$ and where we have integrated down to a length scale $r = e^u \epsilon$ ($\epsilon $ is a UV cutoff).  We can now compute
\beq
\partial_u \lambda = 2 g N_c \partial_u g = 2 b N_c^2 g^4 = 2 b \lambda^2
\eeq
which shows that $\lambda$ has a sensible RG flow in the large $N_c$ limit.  Third, we note that the quartic coupling of the fields $g^2$ goes to zero more slowly as a function of $N$ than in the vector model.  This is a clue that the matrix model is more strongly correlated than the vector model.

A particularly special matrix model is provided by $\mathcal{N}=4$ super Yang-Mills theory in $3+1$ dimensions \cite{magoo}.  This theory has gauge group $SU(N_c)$ and contains many matter fields all in the adjoint representation.  To be specific, the theory contains six real scalars and 4 Weyl fermions as well as the gauge field.  The scalars and fermions carry a large symmetry, called an R-symmetry, equivalent to $SU(4)\sim SO(6)$.  This theory is very special because the matter content is just right so that the $\beta$ function for $\lambda$ vanishes for all $\lambda$.  Thus this theory actually describes a fixed line of conformal field theories.  All this is possible thanks to the very large amount of supersymmetry $(\mathcal{N}=4)$ present in the theory.  The most important scaling operators are so-called single trace operators of the form $\text{tr}(O_1 ... O_n)$ where $O_i$ are adjoint charged fields e.g. the scalar, fermions, or field strength.  At weak coupling the dimension of such an operator would be roughly $n$.  Later on we will meet the holographic dual of this theory.

There are lattice models preserving a large amount of supersymmetry which are believed to flow to the $\mathcal{N}=4$ theory in the infrared \cite{susy_N4_lattice}.  Considerable fine-tuning is still required even with the large amount of lattice supersymmetry, but in principle entanglement renormalization can be carried out even for the $\mathcal{N}=4$ theory.

\subsection{Entanglement and entanglement renormalization}

Having introduced conformal symmetry and the idea of a large $N$ limit, we now review entanglement in many-body systems.  Entanglement renormalization illuminates the structure of entanglement in CFTs in a way that is commensurate with scale invariance and suggestive of holography.

\begin{figure}
\includegraphics[width=.45\textwidth]{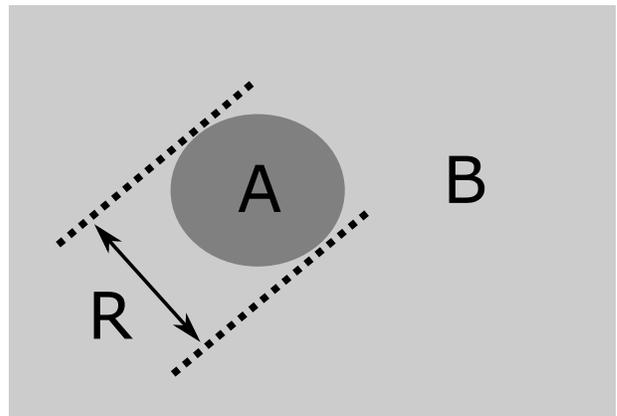}
\caption{Basic setup of entanglement entropy calculations.  The system is divided into two components, here called $A$ and $B$.  The smaller of the two is $A$ which has linear size $R$.  The entanglement entropy $S(A)$ is typically proportional to $|\partial A|$ which in $d=2$ dimensions would be $R$.  This scaling of entanglement with boundary size is called the area law.}
\label{eeAB}
\end{figure}

Entanglement is an important kind of non-classical correlation that appears in composite quantum systems like the one shown in Fig. \ref{eeAB}. A composite system $AB$ is said to be entangled if the the state of the whole system $|\psi_{AB} \rangle$ fails to factorize over the subsystems $|\psi_{AB}\rangle \neq |\psi_A\rangle \otimes | \psi_B \rangle$.  A convenient way to measure entanglement is provided by entanglement entropy or more generally entanglement Renyi entropy.  The Renyi entropy is defined as
\beq
S_n(A) = \frac{1}{1-n} \ln{\left(\text{tr}(\rho_A^n)\right)}
\eeq
where $\rho_A$ is the state of $A$ obtain by tracing out $B$.  If the state of $AB$ is pure then we have $S_n(A) = S_n(B)$ and $S_n(A)$ precisely measures the degree of entanglement between $A$ and $B$.  The entanglement entropy $S(A)$ is obtained in the limit $n\rightarrow 1$ and enjoys various special properties.  A related construction is provided by the mutual information, defined for two regions $A$ and $B$ in a larger system $ABC$ by
\beq
\mathcal{I}(A,B) = S(A) + S(B) - S(AB).
\eeq
Since $S(AB) = 0$ if $AB$ is pure, the mutual information in that case measures entanglement, but more generally the mutual information measures the total amount of correlation between $A$ and $B$.  For example, the mutual information bounds connected correlation functions \cite{mi_bound} of operators in $A$ and $B$
\beq
\mathcal{I}(A,B) \geq \frac{\langle O_A O_B\rangle_c^2 }{||O_A||^2 ||O_B||^2}.
\eeq

To understand the basic scaling structure of entanglement in many-body systems, consider the following scaling argument \cite{mi_qft_bgs,ent_ren_holo}.  Let $r$ denote the length scale of interest in the many-body system and let $A$ be a region of linear size $R$.  Locality suggests that the entanglement between $A$ and $B$ at scale $r$ is proportional to the size of $\partial A$ in units of $r$, namely
\beq
dS \propto \left(\frac{R}{r}\right)^{d-1}.
\eeq
$dS$ is also proportional to the RG measure $dr/r$ and hence we should have
\beq
dS \sim \left(\frac{R}{r}\right)^{d-1} \frac{dr}{r}.
\eeq
Integrating from the UV cutoff $r=\epsilon$ to an IR cutoff $r = \min{(R,\xi_E)}$ we obtain
\beq
S \sim \int_\epsilon^{\min{(R,\xi_E)}} dS = \left(\frac{R}{\epsilon}\right)^{d-1} + ...
\eeq
for $d>1$ and
\beq
S\sim \ln{(\min{(R,\xi_E)}/\epsilon)}
\eeq
in $d=1$.  $\xi_E$ is the length scale beyond which the state has no more entanglement, so we integrate to $\xi_E$ or $R$ depending on whether we first run out of entanglement or run out of degrees of freedom in $A$.  The observation that the entropy is typically proportional to the boundary of $A$ in $d>1$ is called the area law (see Fig. \ref{eeAB}) \cite{arealaw1,arealaw2}.

\begin{figure}
\includegraphics[width=.5\textwidth]{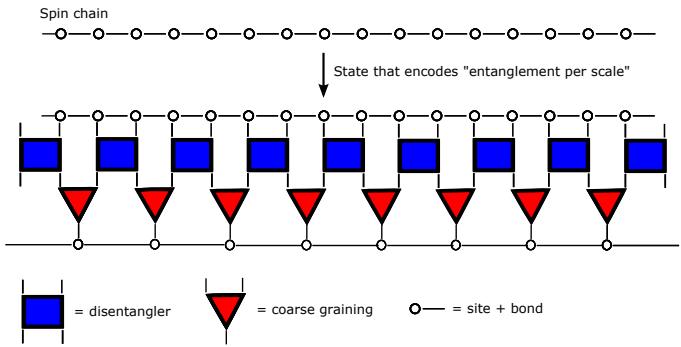}
\caption{Schematic structure of entanglement renormalization with $d=1$ and $k=2$.  The blue squares represent unitaries, called disentanglers, that remove local entanglement.  The red triangles represent isometries, called coarse-grainers, that thin out unentangled degrees of freedom.}
\label{erintro}
\end{figure}

Entanglement renormalization makes the picture above precise and gives a powerful computational tool for many-body systems \cite{vidal_er,ent_ren_holo}.  Entanglement renormalization characterizes the structure of quantum states in terms of two operations, the removal of local entanglement and coarse-graining of degrees of freedom.  We imagine the quantum state of the system at a given length scale $r$.  The degrees of freedom at length $r$ are grouped into blocks to be coarse-grained.  Local entanglement is dealt with using short-ranged unitary transformations.  These act at the edges of blocks to remove short-range correlations between blocks while preserving long-range correlations (blue squares in Fig. \ref{erintro}). In practical terms, the goal may be to reduce the support of the reduced density matrix of a block as much as possible.  Coarse-graining is then implemented by isometries that map the most important states of a block into a renormalized degree of freedom (red triangles in Fig. \ref{erintro}).  Isometries are like unitaries except that they satisfy $W^\dagger W =1$ but not $W W^\dagger =1$.  When the whole system is scale invariant, one expects to be able to remove the same amount of entanglement at each scale and then coarse-grain into equivalent degrees of freedom.  We call the whole network the entanglement renormalization network or simply the MERA network for short.

The use of unitaries and isometries is dictated by the desire to compute correlations efficiently.  Typically, one would compute an expectation value of some local operator $O$ using two copies of the network for the bra and ket in $\langle O \rangle$.  Let us imagine a scheme where $k^d$ sites are to be coarse-grained at each step.  The Hilbert space of a site is assumed to have dimension $\chi$ throughout the network (this can obviously be relaxed).  The state at the UV lattice is related to the state at one lower layer by a layer of disentanglers $U$ and coarse-grainers $W$ as in the schematic expression $|r=\epsilon \rangle = (\otimes U) (\otimes W) | r = k \epsilon \rangle$.  Because unitaries and isometries satisfy $U^\dagger U=1$ and $W^\dagger W=1$ we immediately have
\bea
&& \langle r=\epsilon | r= \epsilon \rangle \cr \nonumber \\
&& = \langle r=k\epsilon | (\otimes W^\dagger) (\otimes U^\dagger) (\otimes U) (\otimes W) | r = k \epsilon \rangle \cr \nonumber \\
&& = \langle r=k\epsilon | r = k \epsilon \rangle.
\eea
We can then repeat the argument for the next layer and thus reduce to calculation of the inner product to the few IR degrees of freedom (in a finite size system).  Similarly, if we want to compute $\langle O \rangle $ then most of the unitaries and isometries will simply cancel to give one, with only a few becoming mixed up with the action of the operator.  In this way, both inner products and local correlations are efficiently computable.

\begin{figure}
\includegraphics[width=.45\textwidth]{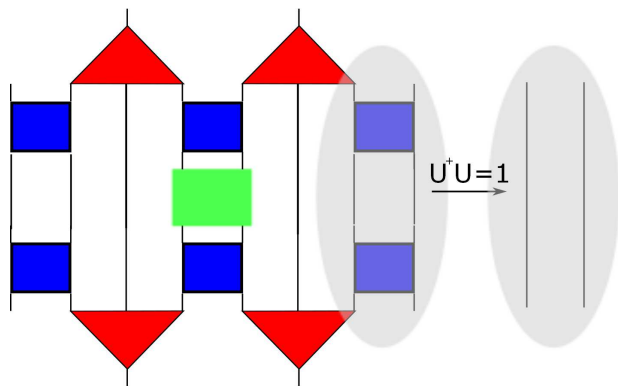}
\caption{An operator (green box) is coarse-grained using one layer of the MERA.  Most of the isometries and unitaries of the layer will give $1$ in this process (substitution in grey shaded region) because of identities like $U^\dagger U = 1$.}
\label{coarsegrain}
\end{figure}

Let us look at the case of correlations in more detail.  Dimensions of operators can be computed using the disentanglers and isometries of the tensor network \cite{mera_opdim}.  We renormalize an operator by applying a layer of of the network to the operator as shown in Fig. \ref{coarsegrain}, and while most of the tensors act trivially, a few will lead to a non-trivial transformation of the operators.  Operators that satisfy $\mathcal{C}(O) = \lambda O$, where $\mathcal{C}(\bullet)$ is the coarse-graining transformation, are called scaling operators because they merely change by a constant factor when renormalized.  When using a $k^d \rightarrow 1$ scheme where $k^d$ sites (in $d$ spatial dimensions) are coarse-grained into one we have $\lambda = k^{-\Delta}$ where $\Delta$ is the dimension of the operator.

\begin{figure}
\includegraphics[width=.5\textwidth]{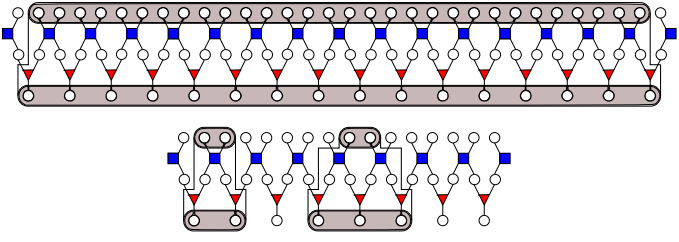}
\caption{The causal cone is the set of all unitaries, isometries, and sites of the network that can affect the state of a region in the UV lattice.  The effective number of sites in a region, and hence the width of the causal cone, shrinks exponentially fast we descend in the network.  This behavior persists until the causal cone width is of order a few lattice spacings where it will fluctuate depending on the details of the scheme.}
\label{ccone}
\end{figure}

\begin{figure}
\includegraphics[width=.5\textwidth]{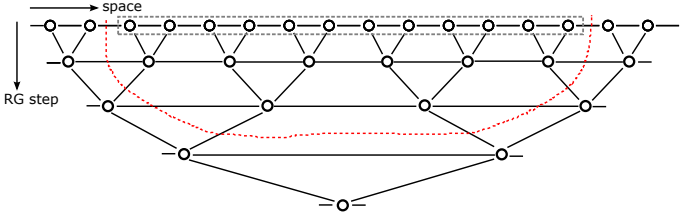}
\caption{Entropy bounds in entanglement renormalization (schematic).  The entropy of a region in the UV lattice (grey boxed in region) is bounded by the number of network bonds that must be cut to isolate it.  The red curve is the corresponding minimal curve which is pierced by the minimal number of bonds.  The length of this curve, suitably defined, or equivalently, the number of bonds cut bounds the entropy.}
\label{erent}
\end{figure}

Another important feature of the entanglement renormalization is the causal cone or RG image \cite{vidal_mera}. The causal cone of a region $A$ in the UV is the set of all sites, unitaries, and isometries in the network that can influence the state of $A$.  As described in Refs. \cite{vidal_er,ent_ren_holo}, the causal cone shrinks exponentially fast as we coarse-grain until it is of order a few lattice sites (the details depend on the precise scheme, see Fig. \ref{ccone}). To compute the entropy of a region $A$, we then start deep down in the causal cone of $A$, when the width of the cone is only a few sites and reverse the RG circuit to reach a more finely-grained scale.  Sites not in the causal cone at the new scale can be traced out.  Since these sites occur along the boundary of the causal cone, the total amount of entropy added is bounded by the size of the boundary of the causal cone.  Roughly speaking, the entropy is bounded by the number of bonds cut times the maximal entropy per site $\ln{(\chi)}$.  We argued extensively in Ref. \cite{ent_ren_holo} that this bound could be interpreted in terms of minimal curves (or surfaces) within the discrete network geometry of the MERA.  For example, Fig. \ref{erent} shows how the number of bonds that must be cut to isolate a region can be greatly reduced by descending into the network instead of isolating the region by cutting it off at the UV scale.  We also argued that correlation functions could similarly be understood in terms of such minimal curves.

The resulting discrete geometry, really a graph with a certain connectivity, approximated smooth AdS space in the case when the lattice model was at a critical point.  This notion of approximation can be made precise in a number of ways, including comparing lengths of corresponding curves in the graph and the smooth space and by comparing the spectrum of the graph and continuous Laplacian \cite{ent_ren_holo}.  Finally, we generalized entanglement renormalization to finite temperature and showed how black hole-like objects appeared in the construction.  Briefly, we considered a circuit $U_{RG}(T)$ whereby the thermal state $\rho(T)$ was represented as
\beq \label{bhcircuit}
\rho(T) = U_{RG}(T) \rho_0 U^\dagger_{RG}(T).
\eeq
Although no longer purely described in terms of entanglement, the disentanglers can still be understood as removing short-range classical and quantum correlation. A black hole is represented by a situation where all correlation is removed and each site of the system sits in a maximally mixed state.  Such a maximally mixed state corresponds to infinite temperature or zero Hamiltonian which is consistent with fact that the black hole horizon is a null surface where the local temperature diverges.

\subsection{Quantum expanders}

We have argued that entanglement renormalization naturally identifies area with entanglement.  Hence to have a classical area in the holographic interpretation of MERA we should find an object that adds a definite amount of entanglement.  Enter quantum expanders.  The definition of a quantum expander begins with the notion of a quantum channel $\mathcal{E}$.  We imagine we have in our control a system $A$ with some state $\rho_A$.  This system $A$ is also in contact with an environment $B$ with which it is initially uncorrelated.  The full state is thus $\rho_{AB} = \rho_A \otimes |0 \rangle \langle 0 |_B$ where $|0 \rangle $ is some reference state.  We are allowed to act with a unitary transformation $V_{AB}$ on $AB$ followed by discarding $B$.  A quantum channel in its simplest form is then the sequence of steps
\bea
&& \rho_A \rightarrow \rho_{AB} \rightarrow V_{AB} \rho_{AB} V_{AB}^\dagger \cr \nonumber \\
&& \rightarrow \text{tr}_B(V_{AB} \rho_{AB} V_{AB}^\dagger) \equiv \mathcal{E}(\rho_A).
\eea
This process is designed to model noise acting on quantum information in the form of a unitary transformation acting on a larger system (to which we do not have access) containing the system of interest.

As a simple example, suppose $A$ and $B$ are two level systems with states $|0\rangle $ and $|1\rangle$ and with $V$ chosen such that
\beq
V_{AB} |00 \rangle = |00\rangle
\eeq
and
\beq
V_{AB} |10 \rangle = |11\rangle.
\eeq
If $\rho_A = |\psi \rangle \langle \psi |$ with $|\psi \rangle = c_0 |0 \rangle + c_1 |1 \rangle$ then
\beq
\mathcal{E}(\rho_A) = |c_0|^2 |0\rangle \langle 0| + |c_1|^2 |1 \rangle \langle 1 |
\eeq
and the state has gone from pure to mixed.  We may interpret this particular quantum channel as decohering or measuring $A$ in the $\{|0\rangle,|1\rangle \}$ basis since it copies the state of the system in this basis to the environment $B$.  Note also that $\rho_A$ was pure while $\mathcal{E}(\rho_A)$ has entropy of $\ln{2}$, so we learn that quantum channels can add entropy (noise) to the system.

It is a standard theorem (Stinespring dilation theorem) that any quantum channel can be written as
\beq
\mathcal{E}(\rho) = \sum_{\alpha=1}^{\alpha_{\text{max}}} M_\alpha \rho M_\alpha^\dagger
\eeq
where the $M_\alpha$ are called Kraus operators and satisfy
\beq
\sum_\alpha M_\alpha^\dagger M_\alpha = 1
\eeq
(proven by tracing over $A$ and using the fact that $V$ is unitary).  Indeed, the operators $M_\alpha$ are nothing but particular sub-matrices of $V$,
\beq
(M_\alpha)_{ij} = (V_{AB})_{i \alpha, j 0},
\eeq
so that
\bea
&& \left(\sum_{\alpha=1}^{\alpha_{\text{max}}} M_\alpha \rho M_\alpha^\dagger\right)_{i l} \cr \nonumber \\
&& = \sum_\alpha (V_{AB})_{i \alpha, j 0} \rho_{jk} (V_{AB}^\dagger)_{k 0, l \alpha} \cr \nonumber \\
&& = \sum_\alpha \langle i \alpha | V_{AB} (\rho \otimes |0 \rangle \langle 0 | ) V_{AB}^\dagger | l \alpha \rangle.
\eea
Note that all we really require to prove this theorem is that the initial state of $AB$ be uncorrelated.

A quantum expander is simply a quantum channel where the matrices $M_\alpha$ satisfy
\beq
M_\alpha = \frac{1}{\sqrt{\alpha_{\text{max}}}} U_\alpha
\eeq
with $U_\alpha$ unitary and where the expander is guaranteed to add entropy to the state \cite{qexpander1,qexpander2}.  The latter condition is obtained by requiring the spectrum of the expander to have a gap.  Eigenoperators of $\mathcal{E}$ are operators such that
\beq
\mathcal{E}(O) = \lambda O
\eeq
and the set of all $\lambda$ is the spectrum of $\mathcal{E}$.  It immediately follows that the operator $1$ is an eigenoperator with eigenvalue $\lambda_1 = 1$.  The spectrum is called gapped, with gap $\delta$, if the next largest eigenvalue satisfies $|\lambda_2| < 1 - \delta$.  It is convenient to choose the quantum expander with $\alpha_{\text{max}}$ even and $U_{\alpha + \alpha_{\text{max}}/2} = U_\alpha^\dagger $ for $\alpha=1,...,\alpha_{\text{max}}/2$.  This insures that the quantum expander is formally a Hermitian operator.

To see what this gap condition has to do with adding entropy, consider the inner product on the space of $\chi \times \chi$ matrices given by
\beq
(M,N) = \text{tr}(M^\dagger N).
\eeq
The normalized operator $1/\sqrt{\chi}$ is an eigenoperator of $\mathcal{E}$ with eigenvalue $1$.  The gap then implies that all other operators are rapidly shrunk the by quantum expander.  Suppose we began with a pure state $\rho_0 = |\psi \rangle \langle \psi |$ and applied $\mathcal{E}$ $n$ times to generate a sequence of states $\{\rho_i \}$ satisfying
\beq
\rho_n = \mathcal{E}(\rho_{n-1}).
\eeq
Since $\rho_0$ is normalized we know that its overlap with $1/\sqrt{\chi}$ is
\beq
(\rho_0,1/\sqrt{\chi}) = \chi^{-1/2}
\eeq
and hence
\beq
\rho_n \sim \frac{1}{\chi} + (1-\delta)^n O_2 + ...
\eeq
This implies that $\rho_n$ approaches the uniformly mixed state exponentially fast and hence we should expect the entropy to increase linearly with $n$.  Of course, how much entropy is actually added depends on the whole structure of the quantum expander.

For another perspective, consider the state $\rho_1$ given by
\beq
\rho_1 = \sum_\alpha \frac{1}{\alpha_{\text{max}}} U_\alpha |\psi \rangle \langle \psi | U_\alpha^\dagger.
\eeq
To the extent that the states $U_\alpha |\psi \rangle $ are orthogonal for different $\alpha$, this expression has $\ln{(\alpha_{\text{max}})}$ more entropy that $\rho_0 = |\psi \rangle \langle \psi |$ (which had none at all).  In the same way, so long as the states $U_\alpha U_{\alpha'} |\psi\rangle $ are orthogonal for $\alpha,\alpha'$ not equal, the state $\rho_2$ has still more entropy, $2\ln{(\alpha_{\text{max}})}$ to be precise.  Thus we see again the linear increase in entropy.

Although we will not directly use the above definition of a quantum expander, these objects are similar to the coarse-graining transformations we consider below.  The main difference will be that while most of the eigenvalues $\lambda$ will be very small, a sparse number of larger eigenvalues will persist even when $\alpha_{\text{max}}$ and $\chi$ are very large.  Nevertheless, these sparse large eigenvalues are not enough to prevent the coarse-graining transformation from removing (or adding) a large and definite amount of entropy from each application.

Typically we will take $\chi \sim e^N$ with $N$ our rough measure of the total number of local degrees of freedom.  It was shown in Ref. \cite{hastings_qexpander_random} that in the context of quantum expanders where the unitaries $U_\alpha$ are drawn randomly from the Haar measure then the second largest eigenvalue is roughly $2/\sqrt{\alpha_{\text{max}}}$ which is very small if $\alpha_{\text{max}} \sim \chi$.  As an aside, in the quantum information context one typically takes $\alpha_{\text{max}}$ to be relatively small (it is analogous to the valence of nodes in a classical expander graph), but our needs are cruder than the full blown small $\alpha_{\text{max}}$ quantum expander.

While generating Haar random unitaries of size $e^N \times e^N$ is exponentially hard in $N$ i.e. we must apply an exponentially large in $N$ number of elementary unitaries, our physical intuition in the area of thermalization tells us that it is not so hard to produce very mixed looking density matrices. Formally, we can invoke a unitary t-design \cite{tdesign} which is simply a measure on the unitary group that agrees with the Haar measure for all polynomials in $U$ of degree less than $t$.  Since these t-designs can be efficiently realized it follows that using generic i.e. Haar random unitaries is not unphysical provided we don't really need arbitrary powers of the unitary.  A recent powerful result is Ref. \cite{random_polydesign} which shows that circuits composed of small random unitaries are approximate polynomial designs.

An extreme form of thermalization known as scrambling has been ascribed to black hole dynamics and large $N$ matrix models, so our considerations are not totally foreign to holography \cite{fast_scramble}.  Scrambling is an extreme form of thermalization where the system is globally thermalized, often at infinite temperature, and looks random even to non-local probes (see Ref. \cite{scramble_quantuminfo} for a discussion). This can be contrasted with local thermalization, hydrodynamic relaxation, and the equilibration of local observables (all of which must happen before full scrambling can occur).

\section{Large $N$ entanglement renormalization}
The generic system we consider has sites supporting a local Hilbert space dimension of roughly $\chi \sim e^N$.  Sites are arranged into a regular graph in $d$ spatial dimensions.  Let $|\Omega \rangle $ be the ground state of a local Hamiltonian $H$ acting on this graph.  Throughout this section we consider entanglement renormalization for $|\Omega\rangle$.  We assume $H$ is translation invariant, $\mathcal{T} H \mathcal{T}^{-1} = H$, where $\mathcal{T}$ is translation by one unit cell.  We also assume that the system is scale invariant e.g. at a critical point.  Later we will relax the second assumption and consider more general kinds of Hamiltonians and states.  The most interesting physics emerges when we consider strongly coupled systems where all $N$ degrees of freedom interact strongly.  We sometimes refer to $|\Omega \rangle $ as the UV state $|UV\rangle$ and the original lattice as the UV lattice.  $k$ denotes the RG scheme i.e. we take $k^d$ sites into $1$ site after coarse-graining. Note that the MERA network is not translation invariant by constructions, so translation invariance is a non-trivial condition on the tensors.

To access the structure of strongly interacting large $N$ theories we assume that the tensors in the network are generic up to constraints imposed by symmetries.  This single assumption underlies our considerations of operator dimensions and area/entanglement relations.  The large $N$ parameter is needed to give an expansion parameter, while the assumption of strong coupling is necessary to assure that the system doesn't decompose into many weakly coupled pieces.  Only a strongly coupled large $N$ system will have disentanglers and coarse-grainers that are truly quasi-random or scrambled.

\subsection{Operator dimensions}

Let us dispense with the trivial case first.  Suppose for a moment we did have $N$ non-interacting copies of a local system on each site.  In this case the disentanglers and coarse-graining transformations would factorize e.g. $U = U_0^{\otimes N}$ for a given disentangler.  We see immediately that the coarse-graining transformation $\mathcal{C}$ will have many eigenvalues of various sizes all throughout its spectrum.  Indeed, if $O$ is a scaling operator in the single copy theory with dimension $\Delta$, then we can trivially form composite operators $\otimes_i O_i$ with dimension $\sum_i \Delta_i$ and hence the spectrum of operator dimensions is quite dense. This situation actually describes the large $N$ vector model \cite{ON_review}, and the weakly coupled matrix model before we project onto gauge invariant states.

To obtain a sparse spectrum of scaling dimensions we must consider more strongly interacting systems.  The coarse-graining operation $\mathcal{C}$ acting on an operator $O$ is
\beq
\mathcal{C}(O) = V^\dagger O V
\eeq
where the operator $V$ is a shallow depth quantum circuit, the product of the all the unitaries $U$ and isometries $W$ in a single layer of the MERA.  This operator is itself an isometry and we note that if the operator $O$ is local then many of the unitaries and isometries in $V$ simply cancel to give factors of $1$. We model strongly interacting systems by assuming the unitaries and isometries strongly mix the Hibert space on which they act.  The similarity to a quantum channel, and in particular quantum expanders, should hopefully be clear.  Now we can ask a sharp question, namely what is the structure of matrix elements of large generic isometries?

Within MERA and more generally, isometries can be understood as unitaries where some of the matrix indices are set to fixed values.  For example, consider a unitary transformation $\tilde{U}: \mathcal{H}^2 \rightarrow \mathcal{H}^2$ ($\text{dim}(\mathcal{H})=\chi$) and define the isometry $\tilde{W}:\mathcal{H}\rightarrow \mathcal{H}^2$ by $\tilde{W}|\psi\rangle = \tilde{U} |\psi \rangle |0\rangle$ where $|0\rangle$ is a reference state.  In this sense an isometry is simply a unitary restricted in its input or output.  Indeed, we can verify that $\tilde{W}^\dagger \tilde{W} = 1_\mathcal{H}$ while $\tilde{W} \tilde{W}^\dagger = \tilde{P}$ where $\tilde{P}$ projects onto the image of $\tilde{W}$.  Now if the unitary $\tilde{U}$ is generic, then its typical matrix elements are of order $1/\chi$.  Furthermore, the action $\tilde{U}^\dagger ( \bullet )\tilde{U}$ will typically turn all operators of the form $(O \otimes 1)$ into the identity.  We consider operators of the form $O \otimes 1$ because they mimic the fact that local operators act on a small subset of the sites.

Here is a formal argument.  Since $\tilde{U}$ is generic, we may as well average over all $\tilde{U}$, thus consider
\beq
\int d\tilde{U}_{Haar} \tilde{U}^\dagger (O\otimes 1) \tilde{U}.
\eeq
$U$ is a $\chi^2 \times \chi^2$ unitary so we have
\beq
\int d\tilde{U} \tilde{U}_{IJ} (\tilde{U}^\dagger)_{KL} = \frac{1}{\chi^2} \delta_{JK} \delta_{IL}.
\eeq
Each of $IJKL$ is a composite index running over $\chi^2$ values, so we can write
\beq
\frac{1}{\chi^2} \sum_{JK} \delta_{JK}\delta_{IL} (O\otimes 1)_{JK} = \frac{tr(O)}{\chi} \delta_{IL}.
\eeq
Here the trace is not the matrix trace above, but the trace over the full Hilbert space.  If $O$ is traceless then this is simply zero, but more generally all operators are immediately reduced to their trace.  Clearly this represents an extreme limit.

To estimate corrections to this picture, consider the operator $O = |n \rangle \langle n|$.  We find that
\bea
&& \langle n 0 | \tilde{U}^\dagger (O \otimes 1) \tilde{U} |n0\rangle \cr \nonumber \\
&& = \sum_k \langle n 0 | \tilde{U}^\dagger |n k \rangle \langle n k | \tilde{U} |n 0\rangle \cr \nonumber \\
&& = \sum_k |\langle n 0 | \tilde{U}^\dagger |n k \rangle|^2 \cr \nonumber \\
&& \sim \frac{1}{\chi}.
\eea
The last line follows because we sum up $\chi$ numbers of rough size $1/\chi^2$.  Hence the operator $O$ is reduced by a factor of $1/\chi$ by the action $U^\dagger (\bullet )U$ and other operators are generated.  Most diagonal elements are now also of order $1/\chi$ while off-diagonal elements are of order $1/\chi^{3/2}$ since phases can cancel in this case e.g. adding $\chi$ random phases gives a number of size $\sqrt{\chi}$.  Similarly, differences in diagonal elements are of order $1/\chi^{3/2}$.  Thus we can say that corrections are of order $1/\chi^p$ for some small number $p$.  This agrees with the perspective from quantum expanders, although again, we are not precisely considering quantum expanders.  In that setting, if $\tilde{U}$ is generic then we would have roughly $\ln{(\alpha_{max})}\sim \ln{(\chi)}$ and hence the expander eigenvalues are of order $1/\chi^p$ according to Ref. \cite{hastings_qexpander_random}.

Repeating this discussion for the actual isometries composing $\mathcal{C}$ for which $\chi \sim e^N$, it follows that the coarse-coarse graining transformation, when acting on local operators, will almost certainly reduce the operator in size by at least a factor of $e^{-N}$.  Since the eigenvalues of the coarse graining operation are $\lambda = k^{-\Delta}$, our estimate shows that most scaling dimensions are very large.  Thus we have two important results.  First, most scaling dimensions are large in a strongly coupled large $N$ theory, and second, at fixed coupling these dimensions increase with the number of degrees of freedom.

In some special cases we can interpolate between the limit of completely random tensors and tensors that decompose into many smaller pieces.  For example, in the $\mathcal{N}=4$ theory the coupling can be varied arbitrarily while preserving conformal invariance.  Thus at fixed $N$ we can smoothly go from the weakly mixed case to the strongly interacting case with an accompanying charge in the spectrum of scaling dimensions.  The fact that a factorized (weakly coupled) large $N$ disentangler can still add or remove a lot of entanglement indicates that the amount of entropy added is partially independent of the spectrum.  Thus in principle there can be multiple scales in the spectrum, set by parameters like $N$ and by the coupling.  In other words, the spectrum is not completely fixed by specifying the entropy added.  We would need a detailed model to determine all the scales, but we emphasize that our basic argument is general: a large $N$ strongly interacting theory will have quasi-random tensors, and such quasi-random tensors lead to a sparse spectrum of dimensions.

If the disentanglers and coarse-grainers in the MERA were truly random then the above result would be the whole story, but this is physically incorrect. Indeed, a state with random tensors is not translation invariant.  Furthermore, there must at least be a stress tensor in the theory which transforms as a scaling operator of low dimension.  This is required by conformal invariance.  Let $h$ be the energy density scaling operator with dimension $\Delta_h$.  In a theory in $d$ dimensions with dynamical exponent $z$ we have $\Delta_h = d+z$ and thus $h$ is a low dimension operator unless $z$ gets large.  If the MERA represents the ground state of the large $N$ model with energy density $h$ then $h$ must be among the eigenoperators of the coarse-graining transformation.  Similarly, if the theory has a bosonic symmetry, the conserved current of that symmetry will also have a small dimension.  Supersymmetry can protect the dimensions of certain operators insuring that they remain low dimension operators.  Importantly, while all these symmetries impose constraints on the structure of the disentanglers and isometries and hence lead to low dimension operators, the generic statement remains that most local operators have a very large dimension under the coarse graining transformation.

Besides these primary operators we also have descendent operators, essentially derivatives of primary operators.  Translation invariance of the MERA guarantees that these operators are also in the spectrum of the coarse-graining superoperator.  Indeed, take a scaling operator $O$ and consider
\beq
\partial O \equiv \frac{O(x)-O(0)}{x} = \frac{\mathcal{T}^x O(0)\mathcal{T}^{-x} -O(0)}{x}
\eeq
where again $\mathcal{T}$ is the translation operator.  The translation invariance of the MERA means that $\mathcal{T} O \mathcal{T}^{-1} $ has the same dimension as $O$ and hence
\bea
&& \mathcal{C}(\partial O) = \frac{k^{-\Delta}O(x/k)-k^{-\Delta} O(0)}{x} \cr \nonumber \\
&& = k^{-\Delta -1} \frac{O(x/k)-O(0)}{x/k} = k^{-\Delta -1} \partial O.
\eea
The last equality implies that $\partial O$ has dimension $\Delta+1$.

Finally, let us address the question of product operators.  Suppose $O_i$ are two low dimension operators, then what can we say about the dimension of $O_1 O_2$?
The operator product expansion in Eq. \ref{ope} is clearly relevant here.  In general, a product of two scaling operators will generate many other scaling operators.  We know, for example, that if $tr(O_1 O_2) \neq 0$ then the identity operator will be among the operators generated.  This follows because the action of $\mathcal{C}$ on an operator can be expressed as $\mathcal{C}(\bullet) = \sum_n \lambda_n O_n (O_n, \bullet)$ (assuming $\mathcal{C}$ is hermitian) where the $O_n$ are normalized eigenoperators.  Large $N$ factorization of correlation functions takes many forms, but here let us simply note that in the context of holography it implies that certain so-called double trace operators, which are essentially squares of single trace operators, have a dimension $\Delta_d = 2 \Delta_s + \mathcal{O}(N^{-p})$ \cite{magoo,condmat_holo_review}.  In our context this follows from the following argument.  Low dimension operators, by definition, largely decouple from the mixing dynamics generated by the coarse-graining transformation.  Thus we expect that these operators are effectively weakly interacting and hence their dimensions should approximately add.  High dimension operators, which are strongly coupled to the mixing dynamics, are not expected to satsify the same property.

\subsection{Entropy and area}

Now we turn to the question of the emergence of area.  We reviewed the arguments of Ref. \cite{ent_ren_holo} that the MERA naturally encodes an emergent holographic geometry in which minimal surfaces provide bounds on entanglement.  Does this bound become sharp in the large $N$ limit? It has already been observed numerically that a single layer of the MERA in $d=1$ tends of add the same amount of entropy at each step of the RG.  This is less surprising when we realize that the $k^d \rightarrow 1$ coarse graining transformation implements a lattice version of the unitary operator $V(D,k) = e^{- i \ln{(k)} D}$ ($D$ is the dilatation generator) so that
\beq
\mathcal{C}(\bullet) \sim V^\dagger(D,k)( \bullet )V(D,k).
\eeq
Remarkably $D$ does indeed have a gap, $\Delta_0$, in its spectrum, so the spectral values $\lambda$ in
\beq
\lambda O = V^\dagger(D,k) O V(D,k)
\eeq
satisfy $\lambda_1 = 1$ ($O_1 = 1$) and $\lambda_2 = k^{-\Delta_0} < 1$.  Thus the continuum operator $V(D,k)$ looks similar to a quantum expander even at finite $N$.  This argument helps explain why the MERA was observed numerically to contribute the same amount of entanglement at every scale.

\begin{figure}
\includegraphics[width=.35\textwidth]{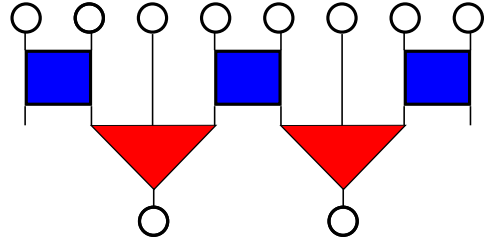}
\caption{Schematic of a $k=3$ MERA in $d=1$.  A site on the lower layer has entropy at most $N$ and the red isometry preserves this when moving from the lower to the upper layer.  The blue disentanglers add roughly $2N$ entropy to the entropy of the three site block which is consistent with an entropy bound for the block of $3N$.}
\label{3to1}
\end{figure}

There is an even stronger statement at large $N$.  If the disentanglers of the MERA are generic in the above sense, then it follows that they will add or remove a definite amount of entanglement when acting on most states.  Let us understand the analogous process for a quantum expander.  Consider again the transformation of density matrices given by
\beq
\mathcal{E} = \frac{1}{\alpha_{max}}\sum_{\alpha=1}^{\alpha_{max}} U_\alpha \rho U^\dagger_\alpha.
\eeq
As we showed above, this transformation arises as unitary transformation on a larger space and the number $\alpha_{max}$ is a measure of the entangling power of that unitary.  We know that if the $U_\alpha$ are random then this transformation is a quantum expander that adds entropy of $\ln{(\alpha_{max})}$ to all but the most mixed states \cite{hastings_qexpander_random}.  If the $\alpha_{max} \sim e^N$ then this transformation will turn every density immediately into the completely mixed state.  However, this is not quite what we want since we're not adding a definite amount of entropy (we're adding just enough to reach the maximally mixed state).

We want to make sure the quantum expander adds a large amount of entropy, of order $N$, but also that it doesn't completely saturate the entropy.  In other words, with $\ln{(\alpha_{max})} = f N$ the quantum expander will add a definite amount of entropy to states whose entropy is not already too high ($S < (1-f) N$).  Of course, the unitaries we are considering in entanglement renormalization do not precisely lead to quantum expanders (for example, the inputs may already be correlated from deeper within the network), but it is still true that a generic unitary acting at the boundary will add an entropy of order $N$ unless the input state is extremely highly entangled already.  Thus we see one of our main results, that generic tensors in the MERA will add definite entropy and have a spectrum of mostly high dimension operators.  Although we again emphasize that the tensors are not totally random since we must have some low dimension operators.

Adding a definite amount of entropy is good, but we also want to see that the tensors correspond to a large local geometry.  To motivate this, consider that the eigenvalue of a scaling operator is a measure of how easily the operator propagates through the network.  Now without modifying the actual network, let us imagine grouping sites in a given layer into supersites of size $\tilde{k}^d$.  To renormalize this supersite into a single site somewhere deeper down requires a larger piece of the network, to be precise we now need $\log_{k}(\tilde{k})$ RG steps.  Using this ``coarse-grained" coarse-graining operator, we see that scaling operators now effectively get reduced by a bigger amount.  But this has a simple interpretation since the operator has really just moved further in the network (multiple layers).  Thus the appearance of small dimensions is naturally interpreted as the existence of an effective large distance in the network.

Even at large $N$ and strong coupling some of the operator dimensions are small, so we know that we haven't just trivially ``coarse-grained" the coarse-graining transformation, but on the other hand, many operator dimensions are large, so it is as if these operators are moving over a much greater distance than the low dimension operators.  This suggests the existence of two lengths scales, one associated with low dimension operators and one with high dimension operators.  Furthermore, if we measure distances in terms of the length scale of the high dimension operators then a single step of the coarse-graining transformation moves operators a large distance.

Another potential concern, noted above, is the appearance of (nearly) maximally mixed state, to which adding entropy is (nearly) impossible.   To address this concern, consider the situation in $d=1$ with $k=3$.  Each site in the network has an entropy of at most $N$, and when we reverse an isometry to reach a less coarse-grained level,
\beq
\rho \rightarrow W \rho W^\dagger,
\eeq
the entropy of the resulting state is identical, see Fig. \ref{3to1}.  However, the resulting state resides in a large Hilbert space of dimension $e^{3N}$ and hence has maximum entropy of $3N$.  When we now act with the disentanglers on edges of the new three site state, we will generate a large amount of entropy since the state of the three sites is far from maximally mixed e.g. it has only $1/k$-th of its maximal entropy (with $k=3$).  Each disentangler will increase the entropy by $N$ still well in line with the total entropy capacity of the region.  More generally, for a $k^d\rightarrow 1 $ scheme in $d$ dimensions, the entropy of a block before the disentanglers act no more than $N$, but since the maximum entropy of the region is $k^d N$, the $2 d k^{d-1}$ disentanglers acting on the boundaries of the block (based on assuming a hyper-cubic block of sides $k$) can still easily add roughly $N$ to the entropy.  Indeed, we only require $k>2d$ or otherwise only add $\frac{k}{2d}N$ entropy per disentangler to be consistent with the bound.

\begin{figure}
\includegraphics[width=.5\textwidth]{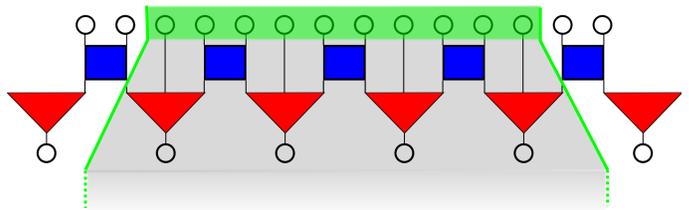}
\caption{Part of the RG image of a block in the UV lattice (green block).  The RG image is the grey shaded region with the green border.  It continues down into the bulk of the network.}
\label{rgimage}
\end{figure}

Let $\delta S_U$ denote the entropy added by the disentangler $U$ to all but the most mixed states.  The total entropy added to a region is then simply $\delta S_U$ times the minimum number of bonds that must be cut to isolate the region.  Thus in the discrete geometry of the MERA network, the entropy of a region is, up to a factor of $\delta S_U \sim N$, simply the minimal number of links piercing the RG image of the region.  The RG image of a region is the causal cone we discussed above, namely all the sites and tensors that can influence the state of the field theory region.  Part of the RG image of region in the UV lattice (green block) is shown in Fig. \ref{rgimage}.  Thus to leading order in $N$ the entropy of a region is completely fixed by the minimal curve needed to isolate it; corrections only occur at order $N^0$.

The geometry of the network is still coarse because of the lattice structure.  However, we argued that a smooth geometry emerges from a combination of the coarse geometry of the tensor network combined with the large tensors within the network.  In other words, each disentangler, say, of the tensor network is really a chunk of a smooth geometry.  One might be worried that these chunks are nevertheless combined in a very non-smooth fashion, but actually this is an illusion.  In a strongly coupled system, the degrees of freedom on any one site are strongly mixed with those of neighboring sites so that the distinction between different sites really loses distinction.  A sharp version of this statement is provided by translation invariance.  Although different sites are formally treated differently by the MERA, translation invariance guarantees that we will get the same result by shifting the entire network by one unit cell.  The resulting network structure looks quite different but still describes the same state.  Thus the sites of the network are really losing meaning deep into the bulk.  It is also interesting to observe that in the context of various topological theories, the bulk network structure of the MERA is explicitly deformable in that new degrees can be exactly introduced into anywhere in the network \cite{mera_topo}.  The topology and connectivity of the network is thus flexible, and the smooth geometry of the bulk is all the more reasonable in this case.  Our freedom to perform the RG in different ways may correspond to bulk diffeomorphisms.

\subsection{Mutual information}

We have now discussed the relation between the form of the spectrum of scaling dimensions and the emergent geometric structure of entanglement.  The structure of the mutual information follows as a natural corollary from the structure of entanglement and correlations.  Consider two regions $A$ and $B$ in the field theory of linear size $R$ separated by a distance $x$.  We ask what happens to these regions when we coarse-grain them using entanglement renormalization.  There are two limits to consider depending on whether $x \gg R $ or $x \ll R$.

\begin{figure}
\includegraphics[width=.45\textwidth]{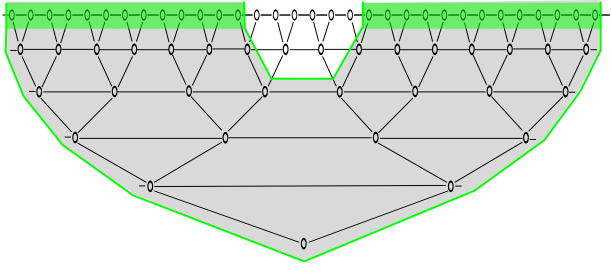}
\caption{RG image or causal cone of two nearby regions.  Notice how the regions merge under coarse-graining before they shrink to the lattice scale.}
\label{mi1dmera}
\end{figure}

If $x \ll R$ then $A$ and $B$ will quickly be merged by the RG flow after a number of steps given by $\log_k{(x/\epsilon)}$ as shown in Fig. \ref{mi1dmera}.  The number $\log_k{(x/\epsilon)}$ is simply how many RG steps to reduce $x$ to $\epsilon$ and where $x$ shrinks by $1/k$ at every step.  This kind of number, which is roughly the RG time needed to shrink a region to the lattice scale, will appear again and again below.  In one dimension the number of bonds cut by the resulting composite region is then
\beq
2 \log_k{((2R+x)/\epsilon)} + 2 \log_k{(x/\epsilon)}
\eeq
where as each region by itself would cut $2 \log_k{(R/\epsilon)}$ bonds.  The mutual information is thus
\bea
&& \frac{\mathcal{I}}{\delta S_U} = 4 \log_k{(R/\epsilon)} \cr \nonumber \\
&& - 2 \log_k{((2R+x)/\epsilon)} + 2 \log_k{(x/\epsilon)}
\eea
or
\beq
\mathcal{I} = 2 \delta S_U \log_k{\left(\frac{R^2}{(2R+x)x}\right)}.
\eeq
which reproduces the results of Ref. \cite{mi_qft_bgs} including the holographic computation.

To give a higher dimensional example consider the case of $d=2$ with $A$ and $B$ two long blocks of length $R$, width $W$, and separated by a distance $x$ (see Fig. \ref{eeholostrip} for an example of one such block).  The number of bonds cut by the merged region is
\beq
\sum_{n=0}^{\log_k{(x/\epsilon)}} \frac{R}{k^n \epsilon} + \sum_{n=0}^{\log_k{((2W+x)/\epsilon)}} \frac{R}{k^n \epsilon}
\eeq
while the number of bonds cut by each region separately is
\beq
\sum_{n=0}^{\log_k{(W/\epsilon)}} \frac{R}{k^n \epsilon}.
\eeq
Hence the mutual information is
\bea
&& \frac{\mathcal{I}}{\delta S_U} = 2 \sum_{n=0}^{\log_k{(W/\epsilon)}} \frac{R}{k^n \epsilon} \cr \nonumber \\
&& - \sum_{n=0}^{\log_k{(x/\epsilon)}} \frac{R}{k^n \epsilon} - \sum_{n=0}^{\log_k{((2W+x)/\epsilon)}} \frac{R}{k^n \epsilon}
\eea
or using
\beq
\sum_{n=0}^N k^{-n} = \frac{1-k^{-N-1}}{1-k^{-1}}
\eeq
we have
\bea
&& \mathcal{I} = \frac{\delta S_U }{1 - k^{-1}}\frac{R}{\epsilon} \left(2 - 2 \frac{1}{k} \frac{\epsilon}{W} - 1 + \frac{1}{k} \frac{\epsilon}{x} - 1 + \frac{1}{k} \frac{\epsilon}{2W+x}  \right) \cr \nonumber \\
&& \sim \frac{\delta S_U }{k - 1} \frac{R}{ x}.
\eea
This is again what was found in Ref. \cite{mi_qft_bgs} on scaling grounds and in holography.  Note also the disappearance of the UV cutoff as in $d=1$.

\begin{figure}
\includegraphics[width=.45\textwidth]{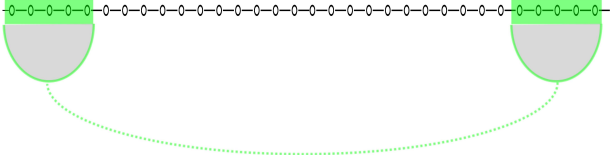}
\caption{Mutual information in the limit $x \gg R$.  The two regions rapidly shrink to the lattice scale and are then only coupled by a few low dimension operators (dotted green line).}
\label{mi1dlong}
\end{figure}

What about the opposite limit where $x \gg R$?  In this case both regions are renormalized to the cutoff scale, taking about $\log_k{(R/\epsilon)}$ RG steps, long before they merge.  Thus all correlations between $A$ and $B$ are due to point like operators originating at the tips of the minimal surfaces (see Fig. \ref{mi1dlong}).  However, we have already seen that the spectrum of operator dimensions is such that only a few low dimension operators exists.  All other operators, even if present, contribute only very tiny corrections, of the order of $(R/x)^{\Delta_{\text{large}}}$.  Thus the mutual information rapidly crosses over from a large order $N$ piece (recall that $\delta S_U \sim N$) to a order one decaying power law at large distances mediated by the few low dimension operators.  Note that this $N$ dependence follows from our result for the large $N$ entropy above since the minimal surfaces for two distant regions will simply be the minimal surfaces for each region separately.  Thus all the $N$ dependence should cancel.

Since the density matrix $\rho_{AB}$ is nearly factorized in the large distance limit, we can compute the mutual information in perturbation theory.  The density matrix is assumed to be $\rho_{AB} = \rho_A \rho_B + \delta \rho$ with $\text{tr}_{AB}(\delta \rho) = 0$. We also define $\delta \rho_A = \text{tr}_B(\delta \rho)$ and similarly for $B$.  The mutual information is
\bea
&& \mathcal{I} = -\text{tr}_A((\rho_A+\delta \rho_A)\ln{(\rho_A +\delta \rho_A)}) \cr \nonumber \\
&& - \text{tr}_B((\rho_B+\delta \rho_B)\ln{(\rho_B +\delta \rho_B)}) \cr \nonumber \\
&& + \text{tr}_{AB}((\rho_A\rho_B +\delta \rho)\ln{(\rho_A \rho_B +\delta \rho)}).
\eea
Expanding to first order in $\delta \rho$ we find
\bea
&& \mathcal{I} = -\text{tr}_A(\rho_A \ln{(\rho_A)} + \delta \rho_A \ln{(\rho_A)} + \rho_A \rho_A^{-1} \delta \rho_A ) \cr \nonumber \\
&& - \text{tr}_B(\rho_B\ln{(\rho_B)} + \delta \rho_B \ln{(\rho_B)} + \rho_B \rho_B^{-1} \delta \rho_B) \cr \nonumber \\
&& + \text{tr}_{AB}(\rho_A \rho_B \ln{(\rho_A \rho_B)} + \delta \rho \ln{(\rho_A \rho_B)} + \delta \rho),
\eea
but this expression actually vanishes.  Thus we have $\mathcal{I} \sim \delta \rho^2$ and since
\beq
\langle O_A O_B \rangle = \text{tr}_{AB} (O_A O_B \delta \rho)
\eeq
for $O_{A,B}$ scaling fields, we expect
\beq
\mathcal{I} \sim \frac{\langle O_A O_B \rangle ^2}{||O_A ||^2 ||O_B||^2}
\eeq
for the correlator of the lowest dimensions operators $O_{A,B}$.  Of course, this is consistent with the bound in Ref. \cite{mi_bound}.  Order $\delta \rho$ terms can appear when considering the Renyi mutual information \cite{mi_renyi_holo}.

Finally, let us compare these strongly coupled results to the situation in a weakly coupled theory.  For the decoupled vector model, the mutual information also decays at large distances like a power law, but the coefficient is not order one but order $N$.  However, the vector model is a non-generic example because the number of primaries with small dimension diverges as $N \rightarrow \infty$.  If the model is weakly coupled but does not have this property, then the spectrum of operator dimensions still won't be sparse, but also won't be diverging.  The picture in Fig. \ref{mi1dlong} is then still valid and although the entanglement geometry is not local at small scales, it still follows that the mutual information has an order one (instead of order $N$) decay at long distances.

\subsection{Correlations, perturbations, and the causal cone}

We now turn to perturbations of the state, for example, in the context of computing correlation functions or studying impurity problems.  An immediate consequence of the causal cone structure of the MERA is that most tensors are not affected when considering a localized perturbation.   For example, local correlation functions may be computed by moving scaling operators through the network using the coarse-graining operator, but the only sites affected are those in the causal cone of the sites where the operators are supported.  For two scaling operators separated by a distance $x$ we must renormalize $\log_k{(x/\epsilon)}$ times to bring the operators adjacent to each other in the network.  Since each operator acquires a factor of $k^{-\Delta}$ after each RG step, the resulting correlation function goes like
\beq
(k^{-2\Delta})^{\log_k{(x/\epsilon)}} \sim \left(\frac{\epsilon}{x}\right)^{2\Delta}
\eeq
as expected.  This is strongly reminiscent of the geodesic approximation in holography, yet the computation here is exact.  A curious feature of the large $N$ limit is that the scaling operators have, in a sense, a lot of room to move within a given site.  They typically represent simple operators, products of a few fields, and hence are not very efficient at moving around within the $e^N$ local states.

\begin{figure}
\includegraphics[width=.48\textwidth]{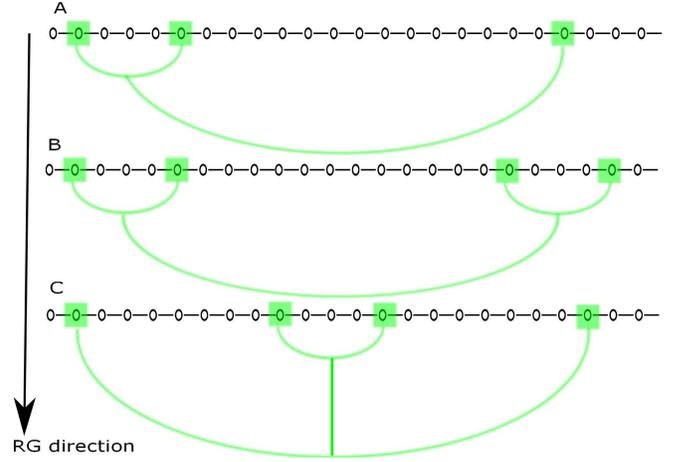}
\caption{Trajectories of renormalized operators in the network.  Panel A shows the three-point function while panels B and C show the four-point function in two different limits.  Note the similarity to Witten diagrams \cite{witten}.}
\label{meracorr}
\end{figure}

As shown in Ref. \cite{ent_ren_holo}, a similar calculation is possible for the three point with the result being a slight generalization of the conformal result (which is appropriate since the MERA can represent scale invariant theories that may not be conformally invariant).  For example, consider three operators $O_1$, $O_2$, and $O_3$ at $x_1$, $x_2$, and $x_3$ and suppose that $|x_1 - x_2| \ll |x_2 - x_3|$.  Then operators $O_1$ and $O_2$ are first merged by the coarse-graining transformation into some operator $O_{12}$ and then this operator and operator $O_3$ are merged much deeper in the network.  We have shown the schematic trajectories of the operators through the network in Fig. \ref{meracorr}.  Conformal invariance implies that $\Delta_{12} = \Delta_3$, since otherwise $\langle O_{12} O_3\rangle $ vanishes (more generally, the smallest dimension operator with non-vanishing correlator with $O_3$ contributes). The coarse-graining transformation then gives the following factors, with $|x_1-x_2| = \ell_<$ and $|x_1-x_3|\sim |x_2-x_3| = \ell_>$,
\beq
\left(\frac{\epsilon}{\ell_<}\right)^{\Delta_1 + \Delta_2} \left(\frac{\epsilon}{\ell_<}\right)^{\Delta_3} \left(\frac{\ell_<}{\ell_>}\right)^{\Delta_{12}} \left(\frac{\ell_<}{\ell_>}\right)^{\Delta_{3}}.
\eeq
If $\Delta_3 = \Delta_{12}$, these factors reduce to
\beq
\langle O_1 O_2 O_3 \rangle \sim \left(\frac{\epsilon}{\ell_<}\right)^{\Delta_1 + \Delta_2 - \Delta_3 } \left(\frac{\epsilon}{\ell_>}\right)^{2 \Delta_3}.
\eeq
which is the usual CFT result
\beq
\langle O_1 O_2 O_3 \rangle \sim \frac{1}{\ell_<^{\Delta_1 + \Delta_2 - \Delta_3} \ell_>^{\Delta_2 + \Delta_3 - \Delta_1} \ell_>^{\Delta_1 + \Delta_3 - \Delta_2}}
\eeq
up to factors of $\epsilon$.  Like the two-point, a geodesic-like structure is evident here as well.

Let us consider then the case of a four-point function which is not completely fixed by conformal invariance.  Let the operators sit at $x_1$, $x_2$, $x_3$, and $x_4$ with $|x_1 - x_2 | = |x_3 - x_3| = \ell_<$, $|x_2 - x_3| = \ell_>$, and $\ell_< \ll \ell_>$.  The operator trajectories are again shown in Fig. \ref{meracorr}.  Operators $O_1$ and $O_2$ and operators $O_3$ and $O_4$ are quickly merged together by the network, but now a new feature appears, since the merged operators $O_{12}$ and $O_{34}$ can take a variety of values and still contribute to the four-point function.  For example, we conformal invariance merely forces $\Delta_{12} = \Delta_{34}$.  Hence the four-point in this limit is given by
\beq
\sum_{\alpha} \left(\frac{\epsilon}{\ell_<}\right)^{\Delta_1+\Delta_2} \left(\frac{\epsilon}{\ell_<}\right)^{\Delta_3+\Delta_4} \left(\frac{\ell_<}{\ell_>}\right)^{2 \Delta_\alpha} c^{\alpha}_{12} c_{34}^{\alpha},
\eeq
where the sum over $\alpha$ is over all operators and the OPE coefficients $c_{ij}^k$ the amplitudes of various fused operators to form new primary fields (see Eq. \ref{ope}).  This sum depends on the full operator content of the CFT and is how effectively how the four-point function becomes non-trivial once we relax the condition of $\ell_> \gg \ell_<$.  We can of course also consider other kinds of limits, for example, if $|x_2 - x_3| = \ell_<$ and $|x_1 - x_2| = |x_3 - x_4| =\ell_>$ then we would obtain a similar sum with a factor of $c_{23}^\alpha c_{14}^\alpha$.  Comparing these forms in different limits leads to what is known as crossing symmetry in CFT; it is a non-trivial constraint on the dimensions and OPE coefficients on the theory.

A general theme of all these computations is the simplifications that occur due to the causal cone structure.  Most of the data in the tensor network is not necessary to compute correlation functions of a few fields or even the entropy of a region.  Thus we have a notion of RG causality. This kind of behavior, where, say, an impurity only affects the network locally, has been observed in the context of boundary critical phenomena in the MERA \cite{mera_bcft1}.  A similar sort of locality has been observed in the context of holographic duals of boundary CFTs \cite{holo_bcft}.  These considerations are also very reminiscent of the ideas in Refs. \cite{gravdual_dmatrix,lightsheet_holo} where an attempt was made to identify bulk regions in a holographic dual that completely encoded the observables of a given field theory region.  It would be very interesting to carry these comparisons further, and to perhaps further study the locality of the bulk RG flow in different contexts e.g. holographic renormalization and Wilsonian renormalization \cite{holo_wilson_rg}.

\subsection{Other geometries}
Having discussed extensively the structure of scale invariant situations, we can obviously consider other kinds of geometries.  The simplest kind of confining geometry is simply one in which the disentanglers successfully remove all entanglement after a certain length scale.  As discussed in Ref. \cite{ent_ren_holo}, the holographic tensor network geometry should be understood as being smoothly capped off in this case.  Correlation functions are identically zero beyond a certain length scale, a hard version of the expected exponential decay.

A more subtle possibility is a situation in which most of the large $N$ degrees of freedom are confined, but a few degrees of freedom remain at low energy.  For example, in the context of the gauged matrix model, the gauge theory could confine at low energies thus removing the large $N$ number of degrees of freedom.  However, a few low lying modes could remain because of symmetry breaking or some other phenomena. In this case our large $N$ considerations only apply above the confinement scale while below it we recover a MERA description with a small number of local degrees of freedom.  This phase of the large $N$ theory could be described as ``non-geometric" although we still have the discrete geometric structure of the tensor network even at small $N$.

We can also imagine the effective dimensionality of space changing.  For example, in Ref. \cite{highe_ads3} it was shown how by turning on a magnetic field one could start with a homogeneous $3+1$ dimensional system and then flow at low energies to a system of many decoupled $1+1$ dimensional systems with a logarithmic violation of the area law.  The holographic version of this story was worked out earlier in the form of a solution that interpolates between $\text{AdS}_5$ and $\text{AdS}_3 \times R^2$.  We can easily implement the same kind of RG flow in the MERA.  As we descend in the network the disentanglers between different one dimensional chains will smoothly go to one as the chains decouple. The chains will eventually totally decouple if all entanglement between degrees of freedom can be removed.  In this case, sufficiently low energy correlators between different chains will vanish, a counterpart of the exponential vanishing of correlators in the $x$-$y$ plane in the holographic setup.

Finally, let us consider the case of a Fermi surface.  Since the Fermi surface is associated with a logarithmic violation of the area law for entanglement entropy \cite{ee_f1,ee_f2,bgs_f1} it has long been observed that the conventional MERA cannot capture the entropy scaling of a Fermi surface with a constant bond dimension.  Of course, the preceding construction does give a state in $3+1$ dimensions with constant bond dimension and logarithmic scaling of the entropy, but it is highly anisotropic.  A related construction known as branching MERA \cite{mera_branch} can give Fermi surface scaling for the entanglement entropy, but it involves a proliferation of different branches of the MERA at low energies.  To understand the physics of this situation, let us examine the dynamics of the Fermi surface more closely.

\begin{figure}
\includegraphics[width=.48\textwidth]{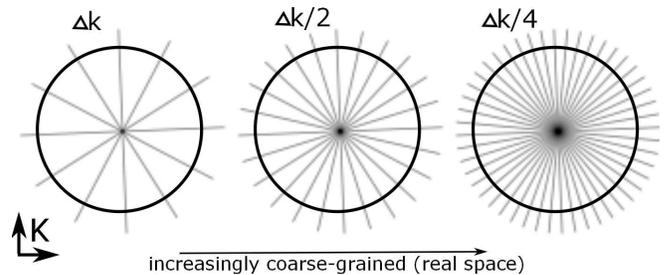}
\caption{Fermi surface coarse-grained into patches of size $\Delta k \sim 1/r$ at scale $r$ in $d=2$.  The number of patches grows in the IR and the Fermi surface contributes more and more to IR sensitive observables.}
\label{fermipatch}
\end{figure}

The Fermi surface of a free fermion system is simply the boundary of the region containing occupied states in momentum space.  In a Lorentz invariant system perturbed by a chemical potential, this surface would typically be a sphere $S^{d-1}$ (in $d$ spatial dimensions) of radius $k_F$ in momentum space.  Now imagine probing this surface with the degrees of freedom at length scale $r$.  Because these degrees of freedom can only probe momentum differences $\delta k\sim 1/r$, the Fermi surface looks coarse to such probes and is effectively divided into $(k_F r)^{d-1}$ patches.  Thus we see that coarse-graining in real space is equivalent to fine-graining in momentum space.  Furthermore, if the effective number of low energy degrees is growing as we renormalize towards the IR, we have a simple model of the area law violation.  If each mode at scale $r$ contributes roughly $(L/r)^{d-1}$ entropy and if there are $(k_F r)^{d-1}$ such modes (see Fig. \ref{fermipatch}), then the entropy should scale like
\beq
S \sim \int_\epsilon^L \frac{dr}{r} \left(\frac{L}{r}\right)^{d-1} \left(k_F r\right)^{d-1} \sim (k_F L)^{d-1} \ln{(L/\epsilon)}.
\eeq

It almost looks as if the Fermi surface system is growing a large extra dimension at low energies.  We can formally define this extra dimension in terms of labels corresponding to points of the Fermi surface.  The fermion operator $\psi(x,t)$ in the free case may be decomposed as
\beq
\psi(x,t) = \int_{FS} d n \, \psi_n(n\cdot x ,t) e^{i k_F n\cdot x}
\eeq
analogous to the decomposition in one dimension.  The fields $\psi_n$ are the slowly varying fields of the low energy description, and we would like to interpret the label $n$, corresponding to a point on a Fermi sphere, as an emergent dimension.  Curiously, the short momentum cutoff is the conventional IR cutoff in real space.  We are free to make this definition, but the crucial question is, does the emergent dimension have a notion of locality which is respected by the physical Hamiltonian?

For free fermions locality is almost trivial since different patches don't interact, but in this case we barely have a smooth space at all, more like a collection of disconnected points.  In the case of the Fermi liquid, the different directions $n$ do interact, but the interactions are totally non-local.  Forward scattering interactions, parameterized by Landau parameters, are not suppressed by distance in momentum space and hence arbitrarily distant points, as measured by distance on the Fermi surface, can interact strongly.  More interesting is the case of strongly coupled Fermi surface gauge field systems \cite{largeN_break_FSGF,nfl_qft1,nfl_qft2}.  In this case the gauge field can only couple nearby patches (and the antipodal patch) efficiently.  Landau-type forward scattering interactions become irrelevant at the new fixed point and a new kind of locality emerges since distant patches are only coupled by irrelevant operators.  This decoupling was used in Ref. \cite{eethermal_cross} to argue for $L \ln{(L)}$ entropy in such phases.  Thus in the strongly coupled system, the label $n$ is much closer to a local coordinate than in the Fermi liquid, yet peculiarities remain, for example, the space may be better understood as a $Z_2$ orbifold since antipodal patches are strongly coupled.  Similarly, there is a curious UV-IR duality at work: long distances in real space correspond to short distances in momentum space and hence short distances in the putative emergent space.  Hence we expect that simultaneous localization is not possible in both spaces, so perhaps there is a connection to non-commutative geometry.

Branching MERA corresponds to the usual MERA except that at each RG step, $b$ new branches are created, with $b=1$ giving the original MERA \cite{mera_branch}.  For example, consider a $k\rightarrow 1$ scheme in $d$ dimensions with $b$ branches created per step.  The size of a given region shrinks after each coarse-graining step by $k$, but $b$ new branches are also created.  Hence the effective entropy that can be generated in a region, which is again proportional to the number of bonds that must be cut to isolate the region, is
\beq
\sum_{n=0}^{\log_k{(L/\epsilon)}} \left(\frac{L}{k^n \epsilon}\right)^{d-1} b^n.
\eeq
Visually, as more an more branches are created, we can literally see a new dimension (or multiple dimensions) growing from the MERA.  To obtain the Fermi surface scaling of entropy we must choose $b=k^{d-1}$ and hence the emergent dimension grows under the RG flow as an effectively $d-1$ dimensional object, exactly like the Fermi surface.  Thus the branching MERA has a natural holographic interpretation.  Just as some systems flow to a reduced effective dimensionality in the IR, other systems can grow new dimensions, and the branching MERA merely reflects this possibility.

Correlation functions also have an interesting structure.  In the branching MERA construction for free fermions, different branches correspond to different pieces of the Fermi surface.  The uncorrelated nature of the branches matches the fact that different points on the Fermi surface are largely uncorrelated.  However, this is not true for Fermi liquids since the Landau forward scattering interactions connect every point to every other point, albeit in a mild way.  If we expect locality to reemerge in the strongly correlated system then the decoupling of patches had better be reflected in the geometry.  From the point of view of geodesics, the growth of the auxiliary sphere representing the Fermi surface makes it very hard for operators to meet deep within the network geometry.  Thus only very nearby operators on the Fermi surface will be correlated.  Much remains to be done if we want to make this picture more precise.

\subsection{Black holes}

There is one additional extremely interesting geometrical feature that we will address.  In Ref. \cite{ent_ren_holo} black hole-like objects were observed in the context of entanglement renormalization at finite temperature.  Strictly speaking, we can no longer speak purely of entanglement since the entropy no longer measures pure entanglement, but the idea of entanglement renormalization still makes sense with the understanding that we are potentially removing both local entanglement and local classical correlations.  The density matrix of the system, assumed to be mixed, is written as
\beq
\rho(T) = U_{RG}(T) \rho_0 U^\dagger_{RG}(T)
\eeq
where $\rho_0$ is a product state.  When $\rho_0$ is a completely mixed state we say the system supports a black hole-like object since the infinite local temperature at a true black hole horizon would produce such a completely mixed state.  An alternate possibility, especially relevant for gapped systems or scale invariant systems on compact spaces, is that the RG flow could stop before a completely mixed state is reached.  In such a case the low energy description of the state is not expected to be strongly affected by a finite temperature.

In most cases equal time correlations in quantum critical system at finite temperature decay exponentially with a correlation length set by the inverse temperature.  In the context of finite temperature entanglement renormalization this decay is turned into a sharp cutoff with cutoff length set by the same scale. For distances small compared to $1/T$ (we assume $z=1$ here) two operators would merge in the network long before they probe the mixed state deep in the network, but once the two operators are separated by much more than $1/T$, their causal cones will terminate at the final disentangled layer before merging.  In this case the operators are totally uncorrelated, a sharp version of the exponential decay.  Curiously, this again looks like the geodesic approximation in holography where geodesics between two boundary points can, beyond a certain separation, disconnect and fall into the black hole horizon.

\subsection{Vector models}

To conclude this section we return to the question of vector models.  Since we have already said that the spectrum of scaling dimensions is quasi-free in the large $N$ limit, it follows that the disentanglers and isometries of the MERA in this case should not be generic in the sense used above.  As partial proof of this statement is found in the computation of universal terms in the entanglement entropy in $2+1$ dimensions.  The entropy scales like $S \sim \alpha L - \beta$ with $\beta_N = N \beta_{free} + \mathcal{O}(1)$ where $\beta_{free}$ is the universal term for a single non-interacting boson \cite{ee_ON_vect,ftheorem_nosusy}.  Thus as measured by the universal properties of entanglement, just as we anticipated above, the $O(N)$ model is relatively weakly interacting.  The different degrees of freedom may still be mixed, but the interactions must be relatively weak to avoid a strongly mixed regime.  We can show that this is the case within the context of MERA as follows.  The vector model
\beq
\mathcal{L}  = \frac{1}{2}\left((\partial \vec{\phi})^2 - m^2 (\vec{\phi})^2\right) - \lambda ((\vec{\phi})^2)^2.
\eeq
has an RG flow from the unstable $\lambda =0$ point to the $\lambda = \lambda_c$ critical point.  This flow takes an RG time of order one and the free fixed point has factorized disentanglers, so the disentanglers at the interacting fixed point are not expected to be strongly mixed.

Vector models are also special in that they have a diverging number of primary fields as $N\rightarrow \infty$.  Other weakly coupled theories without this property are also not expected to have very mixed disentanglers and coarse-grainers and hence should not have a sparse spectrum of operator dimensions.  On the other hand, they may still add a more-or-less definite amount of entropy at each stage of the RG.  The main point is now that we do not have the same separation of scales and hence locality below the scale of the MERA network is not clear.

In contrast, the free limit of $\mathcal{N}=4$ SYM in $3+1$ dimensions is very far from the weakly coupled limit.  We can be quite precise about the situation in this case since there is an entire fixed line parameterized by the 't Hooft coupling that runs from weak to strong coupling.  We see immediately that strong coupling in the gauge theory is very far in parameter space from the weakly coupled limit.  Hence we would expect a long RG flow from the free to the interacting theory and hence potentially very mixed disentanglers.

\section{Comparison to holography}
We will now make a comparison between entanglement renormalization at large $N$ and holographic duality.  Although our considerations are generally independent of dimension, we will often use the $\mathcal{N}=4$ theory in $3+1$ dimensions as a concrete example.  We will first discuss the connection between geometry and entanglement in the gravity theory, then we will turn to operator dimensions before finally discussing the question of bulk locality.

The $\mathcal{N}=4$ theory is dual to type IIB string theory in a spacetime that is asymptotically $\text{AdS}_5 \times S^5$ \cite{magoo}.  The R-symmetry $SO(6)$ is nothing but the symmetry group of $S^5$.  We certainly will not get into the details of string theory, but we will note that the string theory has a number of different kinds of excitations including nearly massless fields known as supergravity fields (including the metric), much heavier string excitations, and other kinds of heavy excitations like D-branes.  The important parameters of the string theory are the string length $\ell_s$, the Planck length $\ell_P$, and the AdS radius $L$.  The AdS radius appears in the metric ($r$ is the holographic direction representing length scale in the field theory)
\beq
ds^2 = \frac{L^2}{r^2} \left(dr^2 -dt^2 + dx^2_d\right)
\eeq
and sets the overall scale.  The gauge theory parameters are related to these length scales via
\beq
\lambda \sim \left(\frac{L}{\ell_s}\right)^4
\eeq
and
\beq
N^2_c \sim \left( \frac{L}{\ell_P}\right)^3.
\eeq
The strong coupling limit, $\lambda \rightarrow \infty$, is therefore equivalent to the supergravity limit where most of the string modes become very heavy and decouple.  Similarly, the large $N$ limit is a classical limit where the Planck length tends to zero and the theory is solvable via saddle point.

We need a few elements of the holographic dictionary to proceed \cite{magoo}.  First, $r=\epsilon$, with $\epsilon$ a small length, is the boundary of the spacetime (the UV of the field theory). Second, bulk scalar fields of mass $m$ are dual to boundary operators of dimension $\Delta$ where
\beq
\Delta (\Delta - (d+1)) = m^2 L^2.
\eeq
There are slight generalizations of this formula that are valid for other kinds of fields e.g. spinors and vectors.  Very heavy bulk fields thus correspond to high dimension operators in the field theory.  Third, at least for Einstein gravity, the Ryu-Takayanagi proposal appears to give the entanglement entropy in the large $N$ limit \cite{holo_ee}.  This proposal says that the entropy of a region $A$ in the field theory is the area in Planck units of the minimal bulk surface anchored to $\partial A$ at $r=\epsilon$.

\subsection{Operator dimensions}
The holographic dictionary relates the mass of bulk fields to the dimension of boundary operators.  Very heavy fields, in units of the AdS radius, correspond to very high dimension operators in the bulk.  If we consider string theory in AdS space, there are at least three different mass scales in the theory.  The lightest fields correspond to the low energy supergravity modes.  They typically have a mass set by the AdS radius $L$ and hence correspond to operators of dimension of order one.  In addition to these supergravity modes we have excited string states.  These states have mass of order $1/\ell_s$ with $\ell_s$ the string length.  Now according to the dictionary above, the 't Hooft coupling in the $\mathcal{N}=4$ is related to $L$ and $\ell_s$ via $\lambda \sim (L/\ell_s)^4$, hence dual operators to these bulk states have dimension
\beq
\Delta \sim L/\ell_s \sim \lambda^{1/4}
\eeq
and hence have a parametrically large dimension in the large $N$ and strongly coupled limit. There are also other heavy states like D-branes with masses related the Planck length and hence also with very large dimensions.

Our considerations above do not match the precise powers of $N$ and $g$ appearing in these formulas, but how could they when we haven't even used an explicit Hamiltonian?  However, we emphasize that we have understood why both the coupling $g$ and the number of states $N$ enter.  Furthermore, there is no obstacle to having a more general structure in the scaling dimensions while still adding entropy of order $N$, we have simply obtained an extreme strong coupling limit.  An alternative model for comparison is provided by the M-theory limit of string theory. In that model the string scale has disappeared so only the Planck length remains.  In this case we may not have a simple weakly coupled description in the phase diagram at large $N$ and most operators will always have large dimension provided only that $N$ is large.

\subsection{Geometry, entanglement, and mutual information}
\begin{figure}
\includegraphics[width=.5\textwidth]{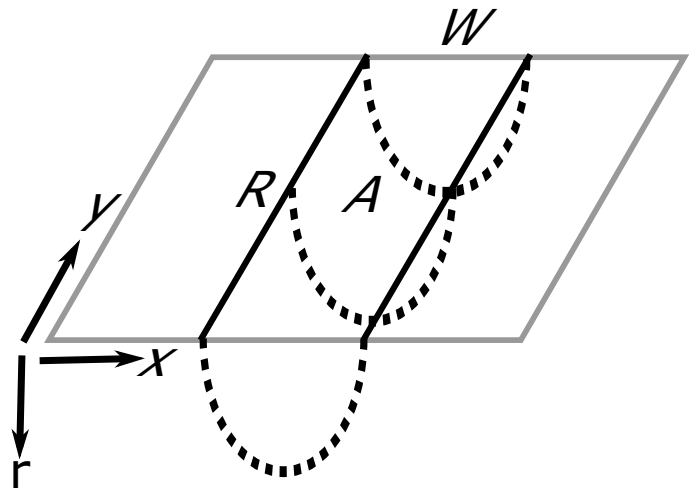}
\caption{Holographic formulation of entanglement.  The field theory region $A$ is a strip of length $R$ and width $W$.  The bulk surface is shown schematically by the dotted lines.  The entropy of the field theory region is the bulk area of the minimal surface anchored to $\partial A$ in Planck units.}
\label{eeholostrip}
\end{figure}

We already established in Ref. \cite{ent_ren_holo} that entanglement renormalization provides a bound on the entropy of a region in terms of a minimal curve in the network geometry.  However, we have argued that at large $N$ each disentangler will add a definite and fixed amount of the entanglement to the state.  Hence the bound we obtained in Ref. \cite{ent_ren_holo} should become tight and we can really identify area and entanglement.  On the holographic side, the Ryu-Takayanagi formula relates entropy to area in the emergent bulk spacetime.  An example of a holographic entropy calculation is shown in Fig. \ref{eeholostrip} where the region $A$ is a strip of length $R$ and width $W$.

Of course, within the context of holographic duality, the RT formula is not the end of the story.  Even if it is correct at large $N$ and $\lambda$ (it has been proven for special geometries, see Ref. \cite{holo_ee_sphere}), it certainly receives corrections at finite $N$ and $\lambda$.  Sometimes the corrections take the form of some kind of modified area formula, but more generally we don't know the answer and have little reason to suspect a connection to the area.  It is true that the higher derivative corrections to black hole entropy, as captured by the Wald formula, will continue to be proportional to the black hole area by construction, this formula cannot be applied more generally since bulk holographic entanglement surfaces as not as simple as black hole horizons\footnote{We thank Rob Myers for explaining this to us.}.  As far as entanglement renormalization is concerned, it appears to be a bulk area law in every case, at least at the level of the coarse network geometry.  The smooth geometry which emerges in the large $N$ limit must come partly from the large $N$ tensors that make up the network.

The structure of holographic mutual information is also qualitatively similar to that obtained in entanglement renormalization.  In both cases we have a relatively sudden transition from a large order $N$ mutual information to a much smaller power decaying mutual information mediated by the few low dimension operators in the spectrum.  Our assumption that the disentanglers are rapidly mixing and hence add a large and definite amount of entropy allows us to give an entanglement renormalization proof of the monogamy of mutual information discussed in Ref. \cite{holo_mi_mono} to leading order in large $N$.  Furthermore, the naive extension of entanglement renormalization to time dependent states should have bearing on the still-open question of monogamy and holographic entanglement in time-dependent backgrounds.  Furthermore, the fact that within entanglement renormalization the large $N$ part of the entropy is always related to some kind area is very suggestive.  Perhaps in general we should minimize some functional of the area to obtain the entanglement entropy holographically at large $N$.  The holographic evidence for this suggestions is very limited (see Ref. \cite{holo_mi_mono} for a brief discussion), so perhaps entanglement renormalization is giving us a strong clue here.  There are still issues, however, since naive proposals, e.g. to use the Wald entropy on non-black hole horizons, appear to fail.

\subsection{Bulk locality}

Finally, let us discuss bulk locality.  We have consistently identified the emergent holographic direction with length scale in the field theory, but this is only half the full story.  More specifically, it appears that this kind of identification is truly generic, and the structure of entanglement renormalization nicely captures this structure even at small $N$.  On the other hand, there are multiple length scales in the bulk, and we can ask about locality with respect to all of them.

In the $\mathcal{N}=4$ theory, there are at least three relevant lengths in the holographic dual, namely the AdS radius $L$, the string length $\ell_s$, and the Planck length $\ell_P$.  Locality on the scale of the AdS radius is related to the coarse RG locality present in the MERA network even at small $N$. At large $N$ each piece of the MERA network adds a large amount of entropy, of order $N$, and hence corresponds on the holographic side to a large area in Planck units, in fact a bulk area of order $L^d$.  This makes sense since we identify $N \sim (L/\ell_P)^3$ ($d=3$) as the number of degrees of freedom in the $\mathcal{N}=4$ theory.  Thus to see bulk locality on the scale of $\ell_s$ or $\ell_P$ we must look within the tensors of the MERA network.  Indeed, while the large $N$ entanglement entropy of the vector model was $N$ times the free result, the large $N$ entanglement of the matrix model does not have this property in general.  Thus the tensors of the tensor network, which control this quantity, are indeed more local in the matrix model since they are effectively sensitive to the metric in Planck units.

Remarkably, we already have partial evidence that there is an emergent local space within these tensors.  This evidence comes from the spectrum of operator dimensions which we argued was sparse.  It was shown in Ref. \cite{holo_from_cft}, by considering a kind of bulk scattering experiment, that just such a sparse spectrum was necessary for bulk locality beyond the AdS radius.  Locality at the scale of the AdS radius was called coarse holography while locality at shorter scales was called sharp holography.  To us it looks like the MERA naturally provides a coarse holographic dual for any theory, where as a sharp holographic dual only emerges in the large $N$ and strongly interacting limits.  In contrast, the higher spin theories which have been suggested to be dual to vector models (specifically, the singlet sector) contain fields of low dimension and arbitrarily high spin.  Based on our considerations here, these fields could lead to non-local behavior up to the scale of the AdS radius, but beyond that they should become effectively local to be consistent with very general and coarse features of renormalization.

It is intriguing to think of the unitaries and isometries as being related to the physics of large $N$ quantum mechanical systems.  In these systems we know that that geometry can emerge from the quantum mechanics of a few large matrices \cite{Mqm_review}.  This is speculation, but we suspect that such models may be used to explain in more detail the origin of a large space \textit{within} a given disentangler in the appropriate strong coupling limit.

\section{Continuous MERA and holography}
We now recast our considerations in the context of the continuous MERA introduced in Ref. \cite{ent_ren_cont}.  The continuous MERA automatically preserves translation invariance and works directly in the continuum but otherwise shares many features with the discrete case.  A major drawback of the continuous MERA is that it is not as computationally useful as the discrete MERA.  The primary advantage of the continuous MERA will be the simple way in which it enables us to define the bulk geometry.  Continuous MERA may thus be a fruitful midway point between full-scale holography and the computational power of the discrete MERA.

As originally formulated, each infinitesimal RG step involves two operations, a disentangling operation and a scaling operation designed to effectively change the number of degrees of freedom.   Both operations are represented by unitary operators, so we can without trouble follow the RG flow in either direction.  Let $u$ denote the RG time starting from the UV and increasing towards the IR.  $u$ is related to our earlier length scale $r$ by $r=e^u \epsilon$.  Deep in the IR, assumed to be unentangled, we introduce entanglement using a transformation
\beq
e^{ -i K(u) \delta u}
\eeq
and then rescale the state using
\beq
e^{ -i D_0 \delta u}.
\eeq
$K$ is an entanglement generating operation and $D_0$ was, in the original work, chosen to be a free dilatation generator.  Both are local operators and are taken to vanish at high energies i.e. they only involve fields below some cutoff.  The choice of $D_0$ presupposes a nearby free fixed point, so more generally we would expect to combine $K$ and $D_0$ into the full dilatation generator $D$ at a fixed point.  If the system undergoes an RG flow then one will have a crossover phenomena where the structure of the dilatation generator changes.  Nevertheless, this is simply a variational ansatz, with variational parameters $K$ and $D_0$, so we need not worry if a precise identification in terms of scaling generators away from fixed points is lacking.

If $|IR\rangle $ denotes the unentangled IR state then our ansatz for the UV state is
\beq
|UV \rangle = V |IR\rangle
\eeq
with
\beq
V = \mathcal{P}_u \left( \exp{\left(- i \int^0_{u_{IR}} du' (K(u') + D_0(u'))\right)} \right)
\eeq
and where $\mathcal{P}_u$ denotes anti-path ordering in $u$.  The conventions here are a little peculiar, but what we are doing is integrating from the IR (large $u$) to the UV (small $u$) and we want the IR parts to hit $|IR \rangle$ first. If the system is near a critical point then we should be able to replace $K+D_0$ with a $u$-independent $D$, the dilatation generator.  In this case we would have $V = e^{- i u_{IR} D}$ except for some modifications in the deep UV or IR.

Local operators inserted in the UV theory at $u=0$ can be moved to different values of $u$ by following the dynamics generated by $D$. Thus we have
\bea
&& O_{UV} |UV\rangle \cr \nonumber \\
&& = O_{UV} e^{- i u D} e^{-i (u_{IR}-u)D} |IR\rangle \cr \nonumber \\
&& = e^{- i u D} O_u e^{-i (u_{IR}-u)D} |IR \rangle,
\eea
where $O_u$ is simply
\beq
O_u =e^{i u D} O_{UV} e^{-i u D}.
\eeq
Now if $O=O_{UV}$ is a scaling operator satisfying
\beq
[D,O(x)]=i \Delta O + i x \partial_x O(x)
\eeq
then
\beq \label{dilatationO}
O_u = e^{-u \Delta } O(x e^{-u}).
\eeq
Note how the inverse transformation is applied to the argument of the field.

If we now insert two scaling operators a distance $x$ apart at $u=0$, they are brought to within $\epsilon$ of each other when $x e^{-u} = \epsilon$.  Each operator will then have been renormalized by a factor of
\beq
e^{-\Delta u} \sim \left(\frac{\epsilon}{x}\right)^\Delta,
\eeq
and since the expectation value of $O_{UV}(\epsilon) O_{UV}(0)$ is independent of $x$, it follows that
\beq
\langle O_{UV}(x) O_{UV}(0)\rangle \sim \frac{1}{x^{2\Delta}}.
\eeq

The discussion of the spectrum of operator dimensions is even more straightforward in the continuous MERA.  The scaling operator $D$, which we suppose is simply a UV cutoff of the continuum dilatation generator (which is anyway not scheme independent), obviously has the properties we want.  It is translation invariant, it has a small number of low dimension states, and most of the states have very large dimension.  Translation invariance requires some explanation.  The dilatation generator is not itself translation invariant as follows from Eq. \ref{cftalg}.  We have $e^{i a P} D e^{- i a P} = D + i a [P,D] = D + a P$, but this result is sensible since we have to choose an origin for $D$.  The crucial point is then that if we apply $D$ to a translation invariant state with $P|\psi\rangle = 0$ then the resulting state is also translation invariant $P D |\psi\rangle = (D P - i P) |\psi \rangle =0$.

The other properties of the dilatation generator are evident from the fact that the operator $D$ is the Hamiltonian of the CFT on a sphere.  The small number of low dimension operators correspond to the small energies of the sphere Hamiltonian while the large number of high dimension operators correspond to a proliferation of excited states at high energy \cite{condmat_holo_review}.  On the holographic side we know the structure of the theory on a sphere from the dual gravitational theory in global AdS.  For example, the Hawking-Page transition demonstrates the distinction between the low and high energy parts of the spectrum.  As always, the details depend on the model, but the fact that $D$ is the integral of a local operator means that our mixing assumption in the discrete setting above is sensible.  Indeed, the scaling superoperator in the lattice MERA is nothing but a regulated version of the unitary flow generated by $D$.

\subsection{Entanglement, geometry, and RG causality}

An important question we need to address is the causality properties of the evolution generated by $D$.  Looking again at the motion of a scaling operator under $D$, we see that it changes position exponentially fast, $O_u(x) = e^{- \Delta u} O(x e^{-u})$.  Thus the dynamics generated by $D$ do not obey a conventional speed limit.  Indeed, looking at the form of $D$ we see that it is explicitly position dependent, so we can understand its properties in terms of a position dependent maximum velocity.  This should be compared to standard Lieb-Robinson bounds showing that all dynamical effects in a lattice model must obey a universal limit, with the speed set by the size of the Hamiltonian.  Thus if the velocity were proportional to $v \sim x$ we would indeed have the observed exponential dynamics since $dx/du = v = x$ implies $x(u) = x(0) e^u$.  So there is a notion of causality, but with a position dependent maximum velocity.  However, we emphasize that the proof in, e.g. Ref. \cite{arealaw_dynamics2}, fails when we add a linearly growing Hamiltonian (basically an extra factorial is generated), so our argument for an exponentially increasing RG light cone is still heuristic.

Nevertheless, the physical picture is compelling.  As operators evolve under the dynamics generated by $D$ they can spread out exponentially fast but not faster. This picture compares favorably with the conventional MERA where operators on distant sites also move together exponentially fast.  Thus it seems that like the discrete case we have a notion of RG causality in continuous MERA.  For example, a perturbation added at a single point should not be able to affect the state outside of the exponentially increasing RG causal cone.  We can compare this with the proposals in Refs. \cite{gravdual_dmatrix,lightsheet_holo} which also tried to identify sub-regions of the bulk geometry that correspond to definite sub-regions in the field theory.  In the context of holographic boundary conformal field theory, the geometry far from the impurity is also untouched in rough agreement with our considerations here \cite{holo_bcft}.

Now, under general conditions, the rate of entropy increase of a region $A$ under time evolution generated by a local Hamiltonian is
\beq
\frac{dS(A)}{du} < \gamma \frac{|\partial A|}{\epsilon^{d-1}}
\eeq
where $\gamma$ is the entanglement generation rate per degree of freedom.  However, it is critical to understand that $\gamma$ depends on the magnitude of the Hamiltonian, and this magnitude is growing when we consider dynamics generated by the dilatation operator.  Thus if $\gamma$ for a region $A$ of size $R$ is effectively $\gamma \sim R$, all the above bound tells us is that
\beq
\frac{dS(A)}{du} < |A|
\eeq
which is rather weak.  Nevertheless, Ref. \cite{ent_ren_cont} assumed that this bound held with a $\gamma$ of order one and then gave a scaling argument, equivalent to our earlier argument, showing that an area law emerges naturally from the RG perspective.  The argument is as follows.  We expect the effective size of $A$ to shrink under the RG time evolution, so the entropy per unit RG time $u$ is 
\beq
dS(A) = \left(\frac{R}{e^u \epsilon}\right)^{d-1} du.
\eeq
When we integrate $dS$ from $u=0$ to $u=\ln{(R/\epsilon)}$ we obtain the area law when $d>1$.

How much entropy is the RG circuit actually adding at every step?  In general, we have called this entropy the entanglement per scale and we identified it as the central charge in $d=1$ CFTs \cite{ent_ren_holo,mi_qft_bgs}.  We start with an unentangled state and then evolve it in time using the dilatation generator.  This scenario is remarkably like that considered in Ref. \cite{ee_1d_quench} which investigated the growth of entanglement after a quantum quench into a CFT from a short-range correlated state.  Ref. \cite{ee_1d_quench} found that the entropy of an interval grew at a constant rate given by
\beq
\frac{dS}{dt} = \frac{\pi c }{6 \epsilon_0}
\eeq
with $\epsilon_0$ some cutoff.  In our case the dilatation operator is dimensionless, so if we make the heuristic replacement $dt/\epsilon_0 \rightarrow du$ then we see that each stage of the RG adds roughly $c$ to the entropy.  Since this holds for any CFT$_{1+1}$ it confirms our earlier argument for constant entropy per RG step even at small $N$ and illustrates that at large $N$ the spectrum of the CFT can be mostly unrelated to the precise amount of entropy added.

There is actually an easy way to obtain a more complete result in our case.  We introduce twist fields $\Psi_n$ that can be used to compute $tr(\rho^n)$ within an $n$-copy replica theory.  To be concrete, the correlator $\langle \Psi_n(0) \Psi_n(L) \rangle_n$ in the $n$-copy theory is
\beq
\langle \Psi_n(0) \Psi_n(L) \rangle_n = \text{tr}\left(\rho_{[0,L]}^n\right).
\eeq
These operators are non-local since they create branch cuts in the $n$-copy theory, yet they behave like conformal primaries of dimension $\Delta_\Psi = \frac{c}{12} \left( n - \frac{1}{n}\right)$ \cite{1d_ee_twist}.  Note that MERA can handle non-local scaling operators \cite{nonlocal_scale_mera}.  Now these operators have the peculiar property that if we take their expectation value in an unentangled pure state, then we get $\langle \Psi_n(0) \Psi_n(L) \rangle_n =1 $ independent of $L$.  This is because $\text{tr}\left(\rho_{[0,L]}^n\right)  =1$ in such an unentangled pure state.

Our task is then to compute the correlator of twist fields in the state $|u \rangle_n = e^{- i u D_n} |IR\rangle_n$ in the $n$-copy theory.  Since $\Psi_n$ is a scaling operator we can use Eq. \ref{dilatationO} to renormalize $\Psi_n$.  If $u < \ln{(L/\epsilon)}$ then the twist fields will be renormalized to the IR before merging, so we obtain the correlator
\bea
&& \langle u | \Psi_n(0) \Psi_n(L) | u \rangle_n \cr \nonumber \\
&& = \langle IR | e^{-u \Delta_\Psi} \Psi_n(0) e^{-u \Delta_\Psi} \Psi_n(L e^{-u}) | IR \rangle_n \cr \nonumber \\
&& = e^{-2 \Delta_\Psi u}.
\eea 
The entropy of the region $L$ in this state is thus
\bea
&& S_n = \frac{1}{1-n} \text{tr}\left(\rho^n_{[0,L]}\right) \cr \nonumber \\
&& = \frac{c}{6}\left(1+ \frac{1}{n}\right) u.
\eea
Taking the $n\rightarrow 1$ limit we have $S=S_1 = cu/3$ and hence $dS/du = c/3$.  This growth with $u$ is cutoff beyond $u=\ln{(L/\epsilon)}$ since the twist fields merge before making it to the IR scale.  Thus going from $|u \rangle $ to $|u + du \rangle$ does indeed add $(c/3) du$ entropy to an interval.

More generally, the above scaling argument and CFT calculations can only take us so far.  They exclude the possibility of a number of low energy degrees of freedom that grows with coarse-graining.  For example, in the case of the Fermi surface it is true that the region shrinks exponentially fast with $u$, but the number of low energy modes also grows exponentially fast with $u$ in precisely the same way.  These two effects then cancel so that
\beq
dS(A) = \left(\frac{R}{e^u \epsilon}\right)^{d-1} (k_F e^{u}\epsilon)^{d-1} du,
\eeq
when integrated, gives a logarithmic correction to the area law.  We also see that in principle even more severe violations of the area law are still consistent with the structure of the evolution generated by the dilatation operator.

With this in mind, let us turn to the interesting proposal in Ref. \cite{cmera_holo_geo} to define a holographic metric based on the continuous MERA.  They take the overlap between states that differ by infinitesimal RG times to define the holographic radial component of the metric.  They showed based on their definition that sensible-looking scale invariant geometries were produced.  One objection is that the large $N$ scaling isn't quite right in their proposal.  This is because the differential overlap should go like $N$ but in the holographic context one expects a different dimension dependent power of $N$ to appear in the metric so that the area in Planck units is ultimately proportional to $N$.  However, this is a minor issue.  Nevertheless, a possible refinement of the proposal in Ref. \cite{cmera_holo_geo} more in line with the original suggestion of Ref. \cite{ent_ren_holo} would be to define local areas in the bulk in terms of differential entanglement generated.

Roughly speaking, we can break the generator of the RG time evolution up into different local pieces.  Each piece acting for an infinitesimal RG time would then give a little piece of bulk area corresponding to $d-1$ field theory directions and the emergent holographic direction.  A more elegant approach is provided by the following construction.  The infinitesimal contribution to the area of a minimal surface anchored at the boundary of $A$ is defined to be proportional to the change in entropy $dS(A)$ where
\beq
dS(A) = S(A,u+du) - S(A,u)
\eeq
where $S(A,u)$ is the entropy of $A$ in the state including degrees of freedom from scales $< u$.  Then the bulk metric would be recovered as an inverse problem in terms of these entropies.  The resulting bulk spatial metric ($r=e^u \epsilon $) can be parameterized in terms of two functions as
\beq
ds^2_{MERA} = f(r) dr^2 + g(r) dx^2_d,
\eeq
and the key problem is to show that the entropy definition is consistent with a geometric computation depending only on these two functions.

We can also use these considerations to address the question of Fermi surface-like entanglement in a holographic setting.  If, as we argued above, the Fermi surface actually gives an equivalent contribution to the entropy at every scale (the decreasing size of $\partial A$ and increasing size of the Fermi surface cancel), then the bulk area element should be
\beq
d A_{bulk} \sim d\Sigma du \sim d \Sigma \frac{dr}{r}
\eeq
where $d\Sigma$ is an infinitesimal element of $\partial A$.  One can simply carry out the continuous MERA optimization procedure for a two dimensional Fermi gas to verify this scaling.  The dilatation generator ends up being a sum of one dimensional dilatation generators for every point of the Fermi surface.

Setting $g(r) = r^{-2}$ as usual, the area element from the metric $ds^2_{MERA}$ is
\beq
d A_{bulk} = \frac{d \Sigma}{r^{d-1}} \sqrt{f(r)} dr
\eeq
which requires $\sqrt{f} = r^{d-2}$.  Writing the metric in a standard form in the putative Fermi surface case we thus find
\beq \label{hyperscaling}
ds^2_{MERA} = \frac{L^2}{r^2}\left( r^{2 (d-1)} dr^2 + dx^2_d\right)
\eeq
Remarkably, just this metric has recently been obtained in the holographic context as a candidate to describe systems with emergent Fermi surfaces \cite{LlogL_holo_ee,hfs_holo_ee}.  The considerations on shape dependence in Ref. \cite{hfs_holo_ee} combined with the results of Ref. \cite{renyi_ff} show that the leading part of entanglement entropy of a Fermi liquid can be consistently interpreted as the area in a higher dimension geometry.

As a technical note, more generally we should take into account the changing shape of $\partial A$, for example, in $d=1$ if the boundary of an interval at scale $r$ is $x(r)$ the area element is
\beq
d A_{bulk} = \sqrt{1 + \left(\frac{dx}{dr}\right)^2} \frac{dr}{r}
\eeq
assuming $f = g = r^{-2}$.  This leads to a slight refinement of our formula above, but this is irrelevant in the Fermi surface case.  Note that our argument is also consistent with a metric that includes an extra $d-1$ dimensional sphere as
\beq \label{expandingFS}
ds^2 = \frac{L^2}{r^2}\left( dr^2 + dx^2_d\right) + (k_F r)^2 ds^2_{S^{d-1}}.
\eeq
In this geometry a minimal surface including all of the sphere will have an area element of
\beq
L^d \frac{d\Sigma}{r^{d-1}} (k_F r)^{d-1} \frac{dr}{r}
\eeq
which also integrates to a logarithmic term.  This should be compared to the branching MERA story above.

Of course, a very similar argument was used to guess the metric in Eq. \ref{hyperscaling} in a purely holographic context (the metric in Eq. \ref{expandingFS} is more novel).  The advantages of our proposal are severalfold.  First, we know by definition that area and geometry are related, while on the holographic side one can question the Ryu-Takayanagi formula.  Second, we know exactly what system we have in mind, an unusual luxury in holography.  Third, we have a precise field theory construction of a putative holographic state.  However, we should emphasize that this is a preliminary proposal, and much additional work needs to be done.

Finally, we cannot help but mention a field theory that potentially combines large $N$ physics with that of a Fermi surface.  This theory is a gauge theory with fermions in the adjoint representation and carrying a conserved charge.  Unlike the vector-like fermion matter content in Refs. \cite{largeN_break_FSGF,nfl_qft2}, this theory has a sensible large $N$ limit with controlled corrections.  By adding a chemical potential for the conserved charge, it becomes possible that this system realizes non-Fermi liquid physics over a wide range of energy scales even if it is eventually unstable.  Similar models in one dimension have recently been been shown to exhibit interesting features in Ref. \cite{strangemetal_1d_cft}.

\subsection{Black holes}

We can also generalize our considerations about black holes to the continuous case.  Consider a thermal state (not normalized) of some conformal field theory given by
\beq
\rho(T) = e^{-\beta H}.
\eeq
We can formally consider the transformed state
\beq
e^{i u D} \rho(T) e^{-i u D} = e^{- \beta e^{-u} H} = \rho(T e^u),
\eeq
where we say ``formally" because the trace of $\rho$ is changed so $e^{i u D}$ is not really a unitary operator in this space.  Nevertheless, a regulated version of this statement is precisely the content of Eq. \ref{bhcircuit}.  Thus starting from some nearly totally thermalized (nearly infinite temperature) state in the IR $\rho_0 = e^{-\beta_0 H}$ with $\beta_0 \approx 0$ we can form the state at a sensible finite temperature by writing
\beq
\rho(T) = e^{-\beta H} = e^{-i u D} \rho_0 e^{i u D}
\eeq \label{bhcmera}
with $u = \ln{(\beta/\beta_0)}$.  In particular, the entanglement geometry of this quantum circuit is completely identical for short distances (distances short compared to $1/T$) to the geometry of the ground state.

The entropy of a region in the field theory will now receive two contributions, one from the entanglement part of the geometry and one from the thermalized degrees of freedom in the deep IR.  Furthermore, the entropy will exhibit an entanglement to thermal crossover as a function of region size and temperature.  Finally, note that the state in Eq. \ref{bhcmera} is only schematic since the unitary $e^{- i u D}$ is only the approximate RG circuit.  In general we have
\beq
\rho(T) = V(T) \rho_0 V^\dagger(T)
\eeq
with $V(T)$ given by $e^{-i u D}$ for most of the RG time but with corrections near the deep UV and IR.  Correlations in this state also follow the general scheme outlined above.  For operators nearer than roughly $1/T$, the two-point function is roughly that of the zero temperature system.  On the other hand, for distances greater than $1/T$ the dilatation dynamics does not bring the two operators together before they run into the decoupled state in the IR.

Let us also compare the total RG flow time.  In our entanglement renormalization proposal the total RG time from UV cutoff to zero IR entanglement is $u = \ln{(\beta/\beta_0)}$.  We convert $\beta$ to a length $r_T = v/T$ using a velocity $v \sim J/\epsilon$ (previously set to one) with $J$ is the energy scale of the cutoff Hamiltonian.  Taking $\beta_0 \sim 1/J$ we get an RG time of $u \sim \ln{(r_T/\epsilon)}$.  On the holographic side, the black hole metric, in the planar AdS$_5$ case, is
\beq
ds^2 = \frac{L^2}{r^2} \left( \frac{1}{f(r)} dr^2 - f(r) dt^2 + dx^2_d\right).
\eeq
The function $f$ is $f(r) = 1 - r^4/r_h^4$ where $r_h$ is the horizon position which scales as $1/T$.  Computing the distance from a UV cutoff $r=\epsilon$ to the horizon $r=r_h \sim r_T$ we find
\beq
\int ds = \int \frac{L}{r\sqrt{f(r)}} dr \sim \ln{(r_T/\epsilon)}
\eeq
in agreement with the entanglement renormalization picture.

\section{Discussion}

In this paper we have considered the problem of entanglement renormalization in strongly coupled large $N$ theories.  We argued that two particular features of these models relevant for holography, namely a sparse spectrum of operator dimensions and the emergence of a classical geometry built from entanglement, emerge naturally from entanglement renormalization.  We also addressed many other features of entanglement renormalization and holography, including generalizations to include Fermi surfaces, the detailed structure of mutual information and correlations, and the generalization to continuous MERA.

Our work has also inspired us to think about several issues in holography proper from a potentially unique direction.  One interesting direction is to consider the role of wormholes in holography.  A classic theorem known as topological censorship \cite{ads_topocensor} forbids the existence of wormholes connecting different asymptotically AdS spaces when the null energy condition is satisfied.  We began thinking about wormholes by asking the question, "If spacetime is entanglement, then is there anything entanglement can't be used to do?"  One simple answer is that entanglement cannot be used to signal in the absence of classical correlations.  Ultimately, this is just a trivial consequence of local dynamics, however, we feel the entanglement perspective is useful.  Thus while spacetime may be composed of entanglement, it should not be possible to use entanglement alone to communicate between different asymptotic boundaries.

Furthermore, since the null energy condition is not expected to be fundamental, e.g. it is violated by quantum and stringy effects, a general principle for a sensible low energy gravity action should be that traversable wormholes are forbidden.  Now in the context of higher spin theories some wormhole geometries have actually been written down, however, it is crucial that in these models the metric is badly gauge non-invariant.  Thus the wormhole may or may not exist in a given gauge.  The challenge here is to formulate the correct notion locality and show that one cannot communicate from one boundary to another.  In theories for which we understand the correct notion of locality, we should simply forbid matter or modifications to gravity that permit wormholes.

It would also be very desirable to formulate our considerations in a more covariant fashion to further bridge the gulf between entanglement renormalization and holography.  Such a covariant formulation would enable us to address exciting dynamical questions including quantum quenches and thermalization.  Of course, we can simply do entanglement renormalization on the time dependent quantum state, but this seems inelegant since time and space are treated very differently.  We do know something about dynamics since the MERA encodes the scaling dimension of the Hamiltonian (and hence effectively the $dt^2$ term in the metric).  Furthermore, using the Heisenberg picture we can compute various time dependent correlation functions starting from an initial state given by MERA, but we would also like to make sense of the Schrodinger picture where one might think of the geometry/tensor network as evolving.  Related to these considerations, we would like to better understand the relationship between entanglement renormalization and other forms of renormalization typically formulated in terms of the partition function or path integral.

Another important question involves symmetry.  We have not focused on symmetry in this work, but it is known how to include symmetry in the tensor network description \cite{tensornet_symmetry}.  This is done with the aid of representation theory and Penrose's spin networks (see Ref. \cite{spinnet_primer} for a gentle introduction).  Those familiar with lattice gauge theory will immediately recognize that the appropriate interpretation of these diagrams is in terms of electric flux \cite{lattgauge_ham,lattgauge_review}.  Thus by tracking the flow of representations down through the MERA we are naturally led to think in terms of emergent electric field lines in the bulk.  This is nothing but the usual statement that a boundary global symmetry is dual to a bulk gauge field.  It would be very interesting to develop this connection further and, for example, look for analogues of Luttinger's theorem \cite{luttingers_theorem} in entanglement renormalization.  There are also intriguing connections between holographic duality and spin networks in loop quantum gravity e.g. Ref. \cite{quantumtet_lqg} which have hardly been explored.

Our discussion here also leads to another natural numerical direction.  One can look for critical theories not be minimizing the ground state of some local Hamiltonian but instead by searching for MERA networks that give rise to translation invariant states.  The idea is that imposing symmetry plus the right structure of entanglement should permit us to write down large classes of critical states without ever needing a parent Hamiltonian.  This is work in progress.  Also, it is not inconceivable that one could just do entanglement renormalization, say in one dimension, for a lattice model that realizes some strongly coupled large $N$ matrix model.  This would be really exciting. We could really fully test holographic duality since we would have the solution on both sides of the duality.

Building on the work of Ref. \cite{ent_ren_holo} and using the large $N$ limit, we have made a strong case for a non-trivial connection between entanglement renormalization and holography.  Of course, the connection could basically stop here, although we think otherwise.  Nevertheless, the situation is positive in any event, if these two beautiful ideas are actually the same thing, great, otherwise they both contain a huge number of unexpected generic features of quantum many-body systems.  Either way our understanding of quantum matter is vastly enhanced.

\textit{Acknowledgements.} I thank Liza Huijse and Subir Sachdev for discussions about entanglement and Fermi surfaces in holography, John McGreevy for many discussions about holography, and Patrick Hayden for insights into quantum information.  I thank Janet Hung and Liza Huijse for valuable feedback on the manuscript.  I am supported by a Simons Fellowship through Harvard University.

\bibliography{space_qexpander}

\end{document}